\documentclass{emulateapj}

\shorttitle{Abundances of Metal-Poor dSph Stars}
\shortauthors{Kirby \& Cohen}
\slugcomment{Accepted to AJ on 2012 September 17}

\begin{document}
\newcommand{\teff}{$T_{\rm{eff}}$}
\newcommand{\mathteff}{T_{\rm eff}}
\newcommand{\logg}{$\log g$}
\newcommand{\mathlogg}{\log g}
\newcommand{\feh}{[Fe/H]}
\newcommand{\mathfeh}{{\rm [Fe/H]}}
\newcommand{\afe}{[$\alpha$/Fe]}
\newcommand{\mathafe}{{\rm [\alpha/Fe]}}
\newcommand{\ah}{[$\alpha$/H]}
\newcommand{\mathah}{{\rm [\alpha/H]}}
\newcommand{\vt}{$v_t$}
\newcommand{\mathvt}{v_t}

\title{Detailed Abundances of Two Very Metal-Poor Stars in Dwarf
  Galaxies\altaffilmark{*}}

\author{Evan~N.~Kirby\altaffilmark{1,2} and
  Judith~G.~Cohen\altaffilmark{1}}

\altaffiltext{*}{Data herein were obtained at the W.~M. Keck
  Observatory, which is operated as a scientific partnership among the
  California Institute of Technology, the University of California,
  and NASA.  The Observatory was made possible by the generous
  financial support of the W.~M. Keck Foundation.}
\altaffiltext{1}{Department of Astronomy, California Institute of
  Technology, 1200 E.\ California Blvd., MC 249-17, Pasadena, CA
  91125} \altaffiltext{2}{Hubble Fellow.}

\keywords{galaxies: dwarf --- galaxies: abundances --- galaxies:
  evolution --- Local Group}

%%%%%%%%%%%%%%%%%%%%%%%%%%%%%%%%%
%%%%%%%%%    ABSTRACT    %%%%%%%%
%%%%%%%%%%%%%%%%%%%%%%%%%%%%%%%%%

\begin{abstract}

The most metal-poor stars in dwarf spheroidal galaxies (dSphs) can
show the nucleosynthetic patterns of one or a few supernovae.  These
supernovae could have zero metallicity, making metal-poor dSph stars
the closest surviving links to Population~III stars.  Metal-poor dSph
stars also help to reveal the formation mechanism of the Milky Way
halo.  We present the detailed abundances from Keck/HIRES spectroscopy
for two very metal-poor stars in two Milky Way dSphs.  One star, in
the Sculptor dSph, has [\ion{Fe}{1}$\rm{/H]} = -2.40$\@.  The other
star, in the Ursa Minor dSph, has [\ion{Fe}{1}$\rm{/H]} = -3.16$\@.
Both stars fall in the previously discovered low-metallicity,
high-\afe\ plateau.  Most abundance ratios of very metal-poor stars in
these two dSphs are largely consistent with very metal-poor halo
stars.  However, the abundances of Na and some $r$-process elements
lie at the lower end of the envelope defined by inner halo stars of
similar metallicity.  We propose that the metallicity dependence of
supernova yields is the cause.  The earliest supernovae in low-mass
dSphs have less gas to pollute than the earliest supernovae in massive
halo progenitors.  As a result, dSph stars at $-3 < {\rm [Fe/H]} < -2$
sample supernovae with ${\rm [Fe/H]} \ll -3$, whereas halo stars in
the same metallicity range sample supernovae with ${\rm [Fe/H]} \sim
-3$\@.  Consequently, enhancements in [Na/Fe] and [$r$/Fe] were deferred
to higher metallicity in dSphs than in the progenitors of the inner
halo.

%The most metal-poor stars in dwarf spheroidal galaxies (dSphs) can show the nucleosynthetic patterns of one or a few supernovae.  These supernovae could have zero metallicity, making metal-poor dSph stars the closest surviving links to Population III stars.  Metal-poor dSph stars also help to reveal the formation mechanism of the Milky Way halo.  We present the detailed abundances from Keck/HIRES spectroscopy for two very metal-poor stars in two Milky Way dSphs.  One star, in the Sculptor dSph, has [Fe I/H] = -2.40.  The other star, in the Ursa Minor dSph, has [Fe I/H] = -3.16.  Both stars fall in the previously discovered low-metallicity, high-[alpha/Fe] plateau.  Most abundance ratios of very metal-poor stars in these two dSphs are largely consistent with very metal-poor halo stars.  However, the abundances of Na and some r-process elements lie at the lower end of the envelope defined by inner halo stars of similar metallicity.  We propose that the metallicity dependence of supernova yields is the cause.  The earliest supernovae in low-mass dSphs have less gas to pollute than the earliest supernovae in massive halo progenitors.  As a result, dSph stars at -3 < [Fe/H] < -2 sample supernovae with [Fe/H] << -3, whereas halo stars in the same metallicity range sample supernovae with [Fe/H] ~ -3.  Consequently, enhancements in [Na/Fe] and [r/Fe] were deferred to higher metallicity in dSphs than in the progenitors of the inner halo.

\end{abstract}

%%%%%%%%%%%%%%%%%%%%%%%%%%%%%%%%%
%%%%%%%%%   SECTION 1   %%%%%%%%%
%%%%%%%%%%%%%%%%%%%%%%%%%%%%%%%%%

\section{Introduction}
\label{sec:intro}

The theory of the hierarchical assembly of the Milky Way (MW) from
smaller structures \citep{sea78,whi78} has enjoyed wide observational
support in the past two decades.  The ongoing disruption of the
Sagittarius dwarf galaxy \citep{iba94} provided dramatic evidence for
presently active hierarchical merging.  \citet{ode01} discovered tidal
tails around the globular cluster Palomar~5, indicating that both
galaxies and clusters participate in the stellar conglomeration.
Perhaps most strikingly, the Sloan Digital Sky Survey has permitted
the discovery of numerous tidal streams \citep{bel06} along many
different lines of sight.  The ubiquity of these merger events makes
it clear that the MW halo is still being formed and that its
constituents are many small objects composed of both stars and dark
matter.

Much of the study of the merging process concerns the precise nature
of the building blocks.  One test of the nature of the halo is its
chemical similarity to smaller objects.  The advent of 8--10~m
telescopes permitted the first detailed chemical analyses of stars in
MW dwarf spheroidal galaxies (dSphs).  \citet{she01,she03} discovered
that the detailed abundance patterns of stars in dSphs did not agree
with stars of similar metallicity in the inner halo.  Therefore, the
surviving dSphs cannot be identical to the primary constituents of the
inner halo.

\citet{rob05} and \citet{fon06a,fon06b} proposed a solution to the
chemical discrepancy.  The inner halo was not built from galaxies like
the surviving dSphs.  Instead, it was built from galaxies closer in
stellar mass and gas content to dwarf irregular galaxies.
Cosmological simulations showed that the inner halo was built very
early by massive satellite galaxies.  In contrast, the surviving dSphs
are small galaxies that formed stars inefficiently.  Also, they have
had over 10~Gyr to alter their stellar populations, including their
chemical compositions.  The surviving dSphs and their siblings are
currently participating in the construction of the outer halo, which
spans a much longer duration than the rapid assembly of the inner
halo.

In the past few years, several studies have attempted to discern
whether galaxies like the surviving dSphs can contribute the most
metal-poor stars to the MW halo.  The search for chemical consistency
was driven by the desire to find a source for the most metal-poor
stars in the halo.  \citet{hel06} claimed that the dSphs are free of
stars with $\mathfeh < -3$ based on line strengths of the infrared
\ion{Ca}{2} triplet.  However, \citet{kir08} discovered such stars in
the MW's ultra-faint dwarf galaxies \citep{sim07}.  Several studies
\citep[e.g.,][]{fre10a,fre10b,sim10,nor10} established the similarity
between these stars' abundances and those of the halo.  A subsequent,
corrected analysis of the \ion{Ca}{2} triplet metallicity
distributions of classical dSphs has uncovered extremely metal-poor
stars \citep{sta10}, and new extremely metal-poor stars have now been
discovered in the classical dSphs using other spectroscopic methods
\citep{kir09,kir10,coh09,coh10,fre10a,taf10}.

Despite the general agreement, some discrepant abundance patterns
persist between the dSphs and the halo.  For example, [Na/Fe] tends to
be lower in the dSphs than in the halo \citep{coh10}, and the
neutron-capture elements are much more enhanced in dSph stars than in
halo stars at $\mathfeh = -1$ \citep{let10}.  As a result of these
minor discrepancies, it is interesting to study chemically peculiar
dSph stars.  \citet[][K10]{kir10} measured the abundances of thousands
of red giants in eight MW dSphs from medium-resolution spectroscopy.
Some of these are both very metal-poor and bright enough for
high-resolution spectroscopic follow-up.  We selected two of the most
interesting stars in this sample.  One of the stars is in the Sculptor
dSph, and, according to \citeauthor*{kir10}, it is magnesium-enhanced
($\rm{[Mg/Fe]} = +0.69 \pm 0.16$).  The other is in the Ursa Minor
dSph, and it is extremely metal-poor ($\mathfeh = -3.62 \pm 0.35$).
In order to verify these interesting abundances and to measure the
abundances of many elements not accessible to \citeauthor*{kir10}'s
study, we obtained Keck/HIRES \citep{vog94} spectra of both stars.
The details of some of our measurement and analysis methods are
somewhat new.  Therefore, we spend much of this article describing our
procedures.

In Section~\ref{sec:obs}, we describe our HIRES observations and data
reduction.  In Section~\ref{sec:ew} we describe our technique for
measuring absorption line strengths and their uncertainties.
Section~\ref{sec:abund} details our method for estimating abundances
and uncertainties from the line strengths.  We interpret our
measurements in Section~\ref{sec:discussion} and summarize our work in
Section~\ref{sec:conclusions}.

%%%%%%%%%%%%%%%%%%%%%%%%%%%%%%%%%
%%%%%%%%%   SECTION 2   %%%%%%%%%
%%%%%%%%%%%%%%%%%%%%%%%%%%%%%%%%%

\section{Observations and Data Reduction}
\label{sec:obs}

\subsection{Target Selection}

Targets were selected from \citeauthor*{kir10}'s catalog of
medium-resolution spectroscopic abundance measurements.  This catalog
contains Fe, Mg, Si, Ca, and Ti abundance measurements from spectral
synthesis of Keck/DEIMOS \citep{fab03} observations.  The sample of
2961 stars is unbiased with respect to metallicity, and it reaches far
down the red giant branch ($V \approx 22$) in eight MW dSphs.
However, the limited spectral range and resolution of DEIMOS restricts
the precision of the abundance measurements and the number of elements
that can be measured.  Therefore, we selected one metal-poor star in
each of two dSphs from the catalog for high-resolution spectroscopic
follow-up to reveal whether the abundance pattern of this star is
consistent with similarly metal-poor stars in the MW halo and other
dSphs.

\subsection{Observations}

\addtocounter{table}{1}

Table~\ref{tab:obs} gives coordinates, photometry, and observational
details for all of our targets.  In addition to Scl~1019417 and
UMi~20103, we included in our study two metal-poor abundance standards
previously observed with HIRES by J.~Cohen: HD~115444 and HD~122563.
Because our equivalent width measurement technique differed from
Cohen's previous work \citep[e.g.,][]{coh03}, we re-analyzed these
spectra to demonstrate the validity of our abundance measurements.
The $V$ magnitude for HD~115444 was taken from the TASS Mark~IV survey
\citep{dro06,ric07}, and the $J$ and $K$ magnitudes were taken from
2MASS \citep{skr06}.  All of the photometry for HD~122563 was taken
from \citet{duc02}.

We configured HIRES optimally for faint, metal-poor red giants.  For
the program star observations, we used the red cross-disperser with a
spectral range of 3927--8362~\AA\@.  The slit width was 1.15'', and
the slit length was 7'', the maximum length that disallows overlap of
the bluest orders.  The spectral resolution depends primarily on the
slit width.  When the seeing is smaller than the slit width, and the
guiding is good, the spectral resolution is smaller than that
corresponding to the projected width of the slit.  Therefore, the
spectral resolution was better for UMi~20103 than for Scl~1019417,
which suffered from the poor seeing at an airmass of $\sim 2$\@.

The spectrograph was configured differently for the abundance
standards than for the program stars because the abundance standards
were observed under programs with different science goals.  The slit
width for HD~115444 was only 0.4''.  The ultraviolet cross-disperser
was used for HD~122563.  The reddest wavelength for this spectrum is
just 5993~\AA, but our line list includes few lines beyond that
wavelength.

The spatial axis of the HIRES CCDs is oversampled, with a plate scale
of 0.12'' per pixel.  Therefore, we binned the detector readout by two
pixels in the spatial axis.  We did not bin the dispersion axis.  The
dispersion is about 0.020~\AA\ per pixel at 5138~\AA\@.

The next section describes the use of DAOSPEC to measure equivalent
widths.  Two products of DAOSPEC are the FWHMs of absorption lines and
the residual spectrum, after all absorption lines have been
subtracted.  We measured the spectral resolving power by calculating
the median of the wavelengths of all lines measured with DAOSPEC
divided by their FWHMs.  We measured the SNR of each spectrum by
calculating the standard deviation of pixels between 5700~\AA\ and
5800~\AA\ in the residual spectrum.  The SNR per pixel is the inverse
of this quantity.  To convert to SNR per resolution element, we
multiplied by the square root of the number of pixels occupied by the
FWHM of weak absorption lines near 5750~\AA\@.  Table~\ref{tab:obs}
lists spectral range, resolving power, and SNR.

The resolution of spectral features is limited by the stellar
macroturbulence.  \citet{gra08} showed that red giants typically have
a macroturbulence of $\sim 5$~km~s$^{-1}$, corresponding to a maximum
resolving power of $R \sim 60\,000$\@.  Therefore, the narrower slit
used for HD~115444 does not ensure that stellar features will appear
at the instrumental resolution, $R = 86\,600$\@.  Indeed, the measured
resolving power for this star is $37\,300$\@.  Taking into account the
spectrograph's line spread function, we estimate that the
macroturbulence for this star is 7~km~s$^{-1}$, consistent with
\citeauthor{gra08}'s (\citeyear{gra08}) relations.

In addition to science exposures, we also obtained exposures of a
thorium-argon arc lamp, a quartz flat lamp, and bright stars to trace
the echelle orders along the detector.

\subsection{Data Reduction}

We reduced the raw frames into one-dimensional spectra using the HIRES
data reduction software
MAKEE\footnote{\url{http://spider.ipac.caltech.edu/staff/tab/makee/}}.
With no user input, this pipeline subtracts the bias level, flat
fields the images, extracts a single one-dimensional spectrum for each
echelle order along the traces determined from the trace star, and
finds the wavelength solution---including the heliocentric
correction---from the arc lamp exposure.  We modified MAKEE slightly
to interpolate over bad columns in the second-generation HIRES CCD
mosaic, installed in 2004.  Some spectral regions were observed on two
or three blue echelle orders.  There were gaps of up to
46~\AA\ between the reddest echelle orders.

We ran MAKEE on each exposure individually.  For the final spectrum,
we added together all of the one-dimensional spectra for a single
object.  MAKEE also provides a $1\sigma$ error spectrum for each
exposure.  We constructed a final error spectrum by adding together
the individual error spectra in quadrature.  MAKEE determined the
wavelength solution from a sixth-order polynomial fit.  DAOSPEC
(Section~\ref{sec:ew}) required a linear wavelength scale.  Therefore,
we linearly rebinned the spectrum for each order.

\begin{figure}[t!]
  \includegraphics[width=\columnwidth]{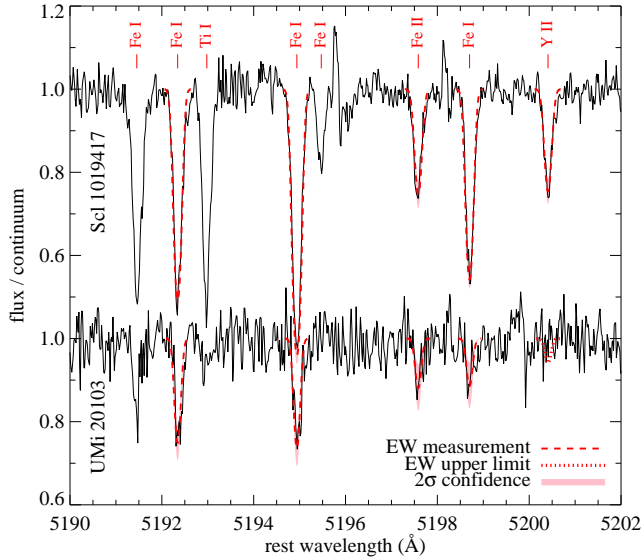}
  \caption{A small region of the HIRES spectra of Scl~1019417 and
    UMi~20103.  Prominent absorption lines are labeled with the
    responsible element and ionization state.  Lines used for
    abundance measurements were fitted with saturated Gaussians
    (dashed red lines).  The light, red shaded regions illustrated the
    $2\sigma$ confidence intervals on the EWs based on line fits to
    different Monte Carlo realizations of the spectra.  The $2\sigma$
    upper limit on the EW for \ion{Y}{2}~$\lambda 5200$ in UMi~20103
    is shown as a dotted red line.\label{fig:specexamples}}
\end{figure}

Figure~\ref{fig:specexamples} shows a small region (5190--5202~\AA) of
the HIRES spectra for the two program stars.  A comparison between the
two spectra shows the lower SNR of UMi~20103.  It is also clear that
UMi~20103 has weaker metal absorption lines than Scl~1019417.  The
weaker lines are due mostly to the higher effective temperature and
partly to the lower metallicity of UMi~20103.

%%%%%%%%%%%%%%%%%%%%%%%%%%%%%%%%%
%%%%%%%%%   SECTION 3   %%%%%%%%%
%%%%%%%%%%%%%%%%%%%%%%%%%%%%%%%%%

\section{Measurement of Equivalent Widths}
\label{sec:ew}

\subsection{DAOSPEC}

We measured equivalent widths (EWs) in each spectrum using DAOSPEC
\citep{ste08}.  This program automatically determines the continuum
shape, absorption line centers, and line strengths.  First, DAOSPEC
iteratively fits and subtracts saturated Gaussian profiles \citep[as
  discussed by][]{ste08} from the observed spectrum.  Then, it fits a
polynomial to the residual flux.  We chose a polynomial order of 20\@.
The program detects additional lines in successive iterations, and it
terminates when the residual spectrum is consistent with flat noise.
This procedure is particularly adept at measuring the EW of each
component of partially blended lines.  The data products are the
continuum shape, residual spectrum, radial velocity, and a line list,
including central wavelengths, line widths, and line strengths.

Each echelle order has its own instrumental response function and
consequently its own continuum shape.  Therefore, we ran DAOSPEC on
each echelle order independently.  We forced the resolving power
($\lambda/\Delta \lambda$) to be constant with wavelength, as is
appropriate for echelle spectrographs.

DAOSPEC cross-correlates the list of detected lines with the user's
list.  Our line list was identical to that of the Keck Pilot Project
\citep{car02,coh03}.  We determined the radial velocity for each star
by examining the difference between the measured line centers and the
predicted line centers from the line list.  Each line gave an
independent measurement of the radial velocity.  In
Table~\ref{tab:obs}, we quote the mean and the error on the mean of
all of the radial velocity measurements for each star.

Although DAOSPEC reported the EW for almost every line in the line
list, some lines required manual intervention.  We built a graphical
user interface, called hiresspec, in IDL to display HIRES spectra and
measure EWs.  Hiresspec shows the linearized spectra of all echelle
orders divided by the continuum polynomial previously determined by
DAOSPEC\@.  Each line from the line list is marked with the responsible
elemental species.  If DAOSPEC successfully measured the line's EW and
FWHM, then hiresspec shows the corresponding saturated Gaussian
profile.  If DAOSPEC failed to measure a line's EW, which happened
rarely, then the user may request that hiresspec fit a saturated
Gaussian to the line.  The FWHM of the line is fixed based on the
spectrum's constant resolving power ($\lambda/\Delta \lambda$).
Hiresspec uses the Levenberg-Marquardt minimization code MPFIT
\citep{mark09} to determine the EW\@.  If a line is not detected above
the noise, the user may request hiresspec to estimate an upper limit.
To do this, hiresspec uses MPFIT to find the EW of saturated Gaussian
that is too strong for the observed spectrum at the $2\sigma$
confidence interval.

\begin{figure}[t!]
  \includegraphics[width=\columnwidth]{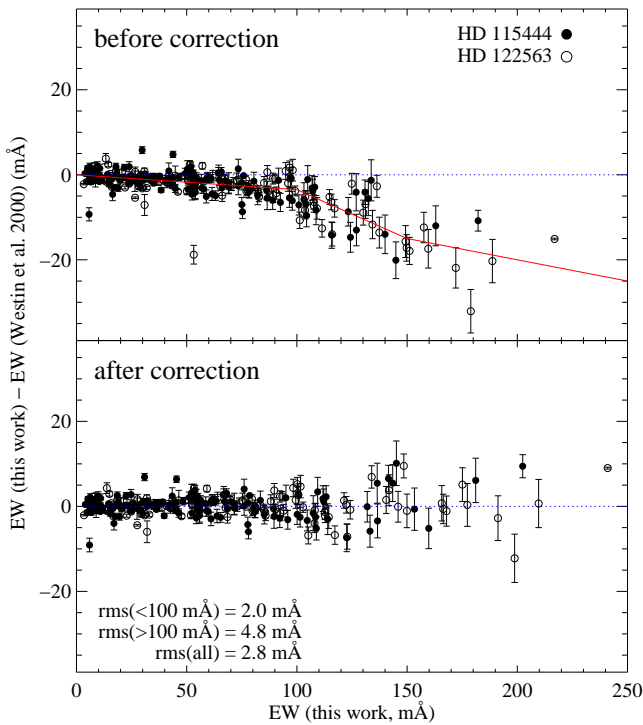}
  \caption{Comparison of the EWs measured here with those of
    \protect\citet{wes00} for the two abundance standards HD~115444
    and HD~122563.  The error bars are calculated from our
    measurements only.  They are the larger of $\delta$EW$_{\rm
      noise}$ or $\delta$EW$_{\rm DAO}$\@.  The top and bottom panels
    show the EWs before and after an empirical correction for the
    imperfect line profile.  The red line in the top panel shows the
    magnitude of the correction as a function of EW.  In the bottom
    panel, the rms is given for three groups: lines that \protect
    \citeauthor{wes00}\ measured to be less than and greater than
    100~m\AA\ and all lines.\label{fig:wes00_ew}}
\end{figure}

Figure~\ref{fig:wes00_ew} compares our EW measurements with those of
\citet{wes00}, who also observed HD~115444 and HD~122563.  Our EWs are
about 3.5\% weaker than those of \citeauthor{wes00}\ at $\rm{EW} <
100$~m\AA\ and about 10\% weaker at $\rm{EW} > 150$~m\AA\@.
\citeauthor{wes00}\ fit Gaussians whereas we fit saturated Gaussians.
The widths of their Gaussians were determined separately for each line
whereas DAOSPEC fixes $\Delta\lambda/\lambda$\@.
\citeauthor{wes00}\ determined the continuum manually for each line
whereas we determined a global continuum for each echelle order.
These are the primary causes for discrepancy between the two sets of
EW measurements.

Our saturated Gaussian fits are not perfect representations for the
HIRES line spread profile.  The profile is complex due to the nature
of the instrument and, for longer exposures, changes in seeing during
the night.  We compared the EWs computed from saturated Gaussians to
EWs computed from direct summation for lines at a variety of strengths
in both HD~115444 and HD~122563.  We found that direct-sum EWs were in
general closer to \citeauthor{wes00}'s EWs than our saturated Gaussian
EWs.  This result suggests that our saturated Gaussian fits
underestimate the true EWs.

Our error estimates (Section~\ref{sec:error}) required an automated
method of determining the spectral continuum and measuring EWs.
Therefore, we chose to use DAOSPEC despite the underestimated EWs.
Instead, we applied an empirical correction to the EWs, depicted in
the top panel of Figure~\ref{fig:wes00_ew}.  For each line with
$\rm{EW} < 100$~m\AA, we raised the EW by 3.5\%.  For each line with
$\rm{EW} > 150$~m\AA, we raised the EW by 10\%.  The correction factor
for lines with $100~{\rm m\AA} < {\rm EW} < 150~{\rm m\AA}$ was
$(0.13({\rm EW/m\AA}) - 9.5)\%$.

\addtocounter{table}{1}

Table~\ref{tab:ew} lists the EW measurements used in the abundance
analysis.  Figure~\ref{fig:specexamples} shows EW measurements for
some metal absorption lines in the two program stars.  All of these
measurements were made with DAOSPEC\@.

Some strong lines are not used in the abundance measurements.  We
eliminated Fe and Ti lines with $\log (\rm{EW}/\lambda) \ga -4.6$ from
the line list.  These lines are roughly on the saturated portion of
the curve of growth.  They are weakly sensitive to abundance and
mostly serve to add noise to the abundance measurements of Fe and Ti,
which have plenty of weaker lines.  Only lines used in the abundance
analysis appear in Table~\ref{tab:ew}.  The upper limits shown in
Table~\ref{tab:ew} are only for those species (element and ionization
state) that do not have a secure measurement of any absorption line.
They are marked with a less-than symbol ($<$).  Table~\ref{tab:ew}
includes only the most restrictive line--the line that yields the
lowest abundance--of any species with an upper limit on its abundance.

\subsection{Error Estimates}
\label{sec:error}

The HIRES spectra of UMi~20103 and Scl~1019417 have moderately low
SNR.  The standard technique of calculating abundance errors, from the
variance of the abundances determined from lines of the same species,
may be inadequate for these spectra.  As a result, we devised a method
for estimating the errors in EWs and abundances caused by spectral
noise.

MAKEE produces an error spectrum, which is mostly random Poisson
noise.  The high spectral resolution makes systematic errors, such as
imprecise subtraction of night sky lines, negligible.  As a result,
the error spectrum may be used to create a different noise realization
of the observed spectrum.  The resampled spectrum ($F_r(\lambda)$) is
equal to the original spectrum with the addition of Gaussian random
noise proportional to the error at each pixel ($\sigma(\lambda)$):

\begin{equation}
F_r(\lambda) = F(\lambda) + R \sigma(\lambda) \label{eq:noise}
\end{equation}

\noindent
where $R$ is a different random number for each pixel, drawn from a
unit normal distribution ($e^{-x^2/2}$).

We resampled each of the four HIRES spectra 100 times.  We ran each
different noise realization through DAOSPEC\@.  DAOSPEC treated each
noise realization independently, with no information from the original
spectrum.  We then used hiresspec in an automated mode to measure EWs
for any lines listed in Table~\ref{tab:ew} that DAOSPEC missed.
Finally, we applied the empirical correction to EWs described in the
previous section.  The final list of EW measurements for each noise
realization contained just as many absorption lines as the original
spectrum.  The final products of the EW measurement process were 101
line lists for each of the four HIRES spectra with EW measurements: EW
measurements for the original spectrum and 100 EW measurements from
noise-added spectra for the same absorption lines.

The EWs quoted in Table~\ref{tab:ew} are the EWs measured from the
original spectrum.  The random errors, $\delta$EW$_{\rm noise}$, are
the standard deviations among the EW measurements from noise-added
spectra.  It may seem that adding noise to the spectrum would inflate
the error estimate.  In other words, the original spectrum already has
noise $\sigma(\lambda)$\@.  Therefore, a noise-added spectrum has
noise $\sqrt{2}\sigma(\lambda)$\@.  However, half of the noise in
every noise-added spectrum is from the same noise realization.  That
component does not change from spectrum to spectrum.  Therefore, the
scatter in EWs among the 100 noise realizations comes only from the
additional component of $\sigma(\lambda)$, not from the unchanging
original noise.

\begin{figure}[t!]
  \includegraphics[width=\columnwidth]{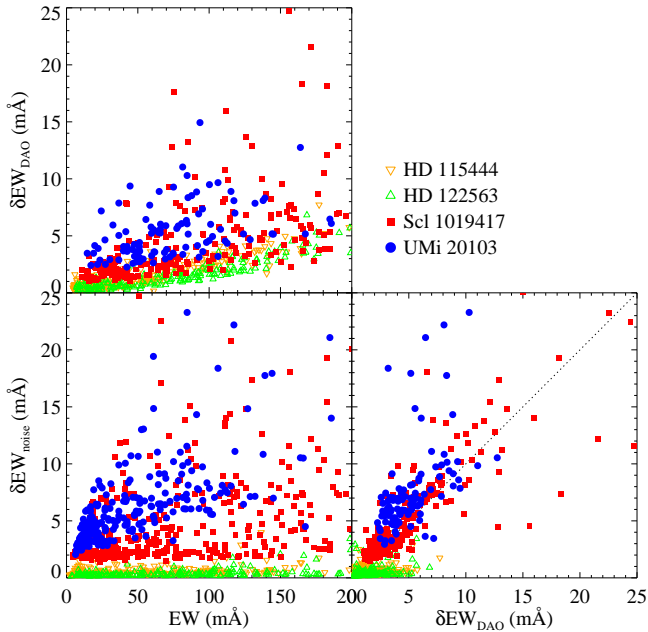}
  \caption{EWs and errors for the four stars observed with HIRES\@.
    Errors are presented both as random measurement uncertainty
    ($\delta$EW$_{\rm noise}$) and the systematic error determined by
    DAOSPEC ($\delta$EW$_{\rm DAO}$).  The dotted line in the lower
    right panel is one-to-one.\label{fig:ewerr}}
\end{figure}

DAOSPEC also reports errors on EWs, which we call $\delta$EW$_{\rm
  DAO}$\@.  This error estimate is more inclusive than
$\delta$EW$_{\rm noise}$\@.  It includes uncertainty on the EW not
only due to spectral noise but also due to systematic errors caused by
blended lines and imprecise continuum placement.
Figure~\ref{fig:ewerr} shows how $\delta$EW$_{\rm noise}$ and
$\delta$EW$_{\rm DAO}$ depend on EW and how the errors relate to each
other.  For the high-SNR spectra of HD~115444 and HD~122563,
$\delta$EW$_{\rm DAO}$ generally exceeds $\delta$EW$_{\rm noise}$,
especially for strong lines.  In these cases, systematic error
dominates the spectral Poisson noise.  The spectra of Scl~1019417 and
UMi~20103 have lower SNRs.  As a result, spectral Poisson noise
dominates the EW errors, and $\delta$EW$_{\rm noise}$ and
$\delta$EW$_{\rm DAO}$ agree well with each other.

%%%%%%%%%%%%%%%%%%%%%%%%%%%%%%%%%
%%%%%%%%%   SECTION 4   %%%%%%%%%
%%%%%%%%%%%%%%%%%%%%%%%%%%%%%%%%%

\section{Measurement of Atmospheric Parameters and Abundances}
\label{sec:abund}

We computed abundances using ATLAS9 \citep{kur93} model atmospheres
and the code
MOOG\footnote{\url{http://www.as.utexas.edu/~chris/moog.html}}
\citep{sne73} with an improved treatment of scattering \citep{sob11}.
This section describes how we determined the atmospheric parameters
(effective temperature, surface gravity, microturbulent velocity,
metallicity, and alpha enhancement) and how we subsequently computed
abundances from EWs and their errors.

\subsection{Atomic Data}
\label{sec:atomicdata}

We used the atomic line list of the Keck Pilot Project \citep{car02}.
\citet{coh03} described the list in detail.  Where available, we
updated the list with version 4.1.1 of the National Institute of
Standards and Technology (NIST) Atomic Spectra Database \citep{ral11}.
We also treated hyperfine structure (see Section~\ref{sec:hfs}) with
\citeauthor{coh04}'s (\citeyear{coh04}) compilation of hyperfine
transitions.  (For simplicity, Table~\ref{tab:ew} lists only one line
for each hyperfine complex.)  Where available, we used damping
constants from \citet{bar00} and \citet{bar05a}.  Otherwise, we used
the damping constants from the Keck Pilot Project's line list.

\subsection{Determination of Atmospheric Parameters}
\label{sec:atm}

The most important atmospheric parameter in the determination of
abundances is the effective temperature (\teff).  This parameter may
be determined from photometry and stellar evolutionary considerations,
or it may be determined directly from the spectrum with no other
input.  \citet{iva01} described in detail the methods and advantages
to calculating \teff\ and surface gravity (\logg) from both the
photometric and spectroscopic methods.  We experimented with
calculating \teff\ from both methods.

\begin{deluxetable*}{lccccccccc}
\tablecolumns{9}
\tablewidth{0pt}
\tablecaption{Model Atmosphere Parameters\label{tab:atmpars}}
\tablehead{ & \multicolumn{4}{c}{Spectroscopic \teff\tablenotemark{a}} & & \multicolumn{4}{c}{Photometric \teff} \\ \cline{2-5} \cline{7-10}
\colhead{Parameter} & \colhead{HD 115444} & \colhead{HD 122563} & \colhead{Scl 1019417} & \colhead{UMi 20103} & & \colhead{HD 115444} & \colhead{HD 122563} & \colhead{Scl 1019417} & \colhead{UMi 20103}}
\startdata
$T_{\rm eff}$ (K)                      & 4750 & 4662 & 4280 & 4799 & & 4771 & 4742 & 4356 & 4855 \\
$\delta_{\rm noise} T_{\rm eff}$ (K)\tablenotemark{b}   &    6 &    6 &   32 &   72 & & \nodata & \nodata & \nodata & \nodata \\[2ex]
$\log g$\tablenotemark{c}              & 1.44 & 1.36 & 0.42 & 1.66 & & 1.44 & 1.36 & 0.42 & 1.66 \\[2ex]
$\xi$ (km~s$^{-1}$)                    & 2.09 & 2.48 & 2.21 & 2.04 & & 2.09 & 2.45 & 2.23 & 2.02 \\
$\delta_{\rm noise} \xi$ (km~s$^{-1}$)\tablenotemark{b} & 0.04 & 0.01 & 0.10 & 0.27 & & 0.04 & 0.02 & 0.13 & 0.22 \\[2ex]
$\rm{[Fe/H]}$                          & $-3.11$ & $-2.89$ & $-2.45$ & $-3.15$ & & $-3.10$ & $-2.81$ & $-2.37$ & $-3.09$ \\
$\delta_{\rm noise} {\rm [Fe/H]}$\tablenotemark{b}      & $0.01$ & $0.01$ & $0.04$ & $0.10$ & & $0.01$ & $0.01$ & $0.03$ & $0.04$ \\[2ex]
$\rm{[\alpha/Fe]}$                     & $+0.57$ & $+0.46$ & $+0.62$ & $+0.37$ & & $+0.57$ & $+0.49$ & $+0.60$ & $+0.37$ \\
$\delta_{\rm noise} {\rm [\alpha/Fe]}$\tablenotemark{b} & $0.01$ & $0.01$ & $0.05$ & $0.09$ & & $0.01$ & $0.00$ & $0.05$ & $0.09$ \\
\enddata
\tablenotetext{a}{These are the values used for the abundance measurements.  The photometric parameters are given for comparison only.}
\tablenotetext{b}{These are error estimates based on spectral Poisson noise only.  Typical total errors, including systematics, for spectroscopically derived values of $T_{\rm eff}$ and $\xi$ are 100~K and 0.2~km~s$^{-1}$.  A typical error for photometrically derived values of $T_{\rm eff}$ is 150~K.}
\tablenotetext{c}{Photometric values of $\log g$ are used.  We do not determine spectroscopic values.}
\end{deluxetable*}

First, we estimated \teff\ and \logg\ from photometry alone.  For the
dSph stars, we corrected the $V$ and $I$ magnitudes for extinction
from \citeauthor{sch98}'s (\citeyear{sch98}) dust maps: $E(V-I) =
0.029$ for Sculptor and $0.042$ for Ursa Minor.  From the $I_0$
magnitude and $(V-I)_0$ color, we calculated \teff\ and \logg\ from
12~Gyr Yonsei-Yale isochrones \citep{dem04} with ${\rm [Fe/H]} = -2.5$
and ${\rm [\alpha/Fe]} = +0.3$\@.  We adopted distance moduli of
$(m-M)_0 = 19.67 \pm 0.12$ for Sculptor \citep{pie08} and $19.18 \pm
0.12$ for Ursa Minor \citep{mig99}.  For the bright abundance
standards, distances are unavailable.  Therefore, we followed the
procedure of \citet{coh02} to determine atmospheric parameters.  Both
\teff\ and \logg\ were determined iteratively from a combination of
color-temperature relations and Yonsei-Yale isochrones.  We corrected
the $V$ magnitudes for extinction \citep{sch98}, and we calculated
$V-J$ and $V-K$ color temperatures \citep{hou00} assuming an initial
guess of $\mathfeh = -2.8$ and $\mathlogg = 1.4$\@.  From the average of
these two temperatures, we computed \logg\ from 12~Gyr, alpha-enhanced
Yonsei-Yale isochrones.  Then, we recalculated color temperatures with
this estimate of \logg.  We repeated this process until \teff\ and
\logg\ converged.  The right half of Table~\ref{tab:atmpars}, under
the heading ``Photometric \teff,'' gives the photometric values of
\teff\ and \logg\ for all four stars.

From \logg, we computed an initial guess for the microturbulent
velocity ($\xi$) from Equation~2 of \citet{kir09}.  We also made an
initial guess at \feh\ and \afe, defined as the average of the
available [Mg/Fe], [Si/Fe], [Ca/Fe], and [Ti/Fe] measurements.  We
took these measurements from \citeauthor*{kir10} for the dSph stars
and from \citet{wes00} for the abundance standards.  From the five
atmospheric parameters (\teff, \logg, $\xi$, \feh\, and \afe), we
interpolated an ATLAS9 model atmosphere from \citeauthor{kir11a}'s
(\citeyear{kir11a}) grid.

With the line list and the model atmosphere, we used MOOG to compute
abundances, $\epsilon$\footnote{We use the notation $\epsilon({\rm X})
  = 12 + \log \frac{n({\rm X})}{n({\rm H})}$, where $n({\rm X})$ is
  the photospheric number density of element X.}, for each line of Mg,
Si, Ca, Ti, and Fe.  We then calculated \feh\ as the average of the
abundances from all of the Fe lines, regardless of ionization state.
We also calculated [Mg/Fe], [Si/Fe], and [Ca/Fe], and we averaged them
to obtain \afe.  These new quantities were used to interpolate a new
ATLAS9 atmosphere and compute new abundances.  We then measured six
quantities, $m_i$, from the abundances derived from Fe and Ti lines:

\begin{enumerate}
\item $m_1$: The slope of abundance with excitation potential (EP),
  $d(\log \epsilon)/dEP$, for \ion{Fe}{1} lines.  This parameter is
  most affected by \teff.

\item $m_2$: The slope of abundance with reduced width, $d(\log
  \epsilon)/d(\log ({\rm EW}/\lambda))$, for \ion{Fe}{1} lines.  This
  parameter is most affect by $\xi$.

\item $m_3$: The difference between the average \ion{Fe}{1} abundance
  and the average \ion{Fe}{2} abundance.  This parameter is most
  affected by \logg.

\item $m_4$, $m_5$, and $m_6$: The same three quantities for Ti lines.
\end{enumerate}

\noindent The slopes were computed from least-squares linear
regressions.  Jackknife errors, $\delta m_i$, were calculated for all
six quantities.  In general, the errors from the Fe lines were a
factor of several smaller than from the less numerous Ti lines.  We
calculated a goodness-of-fit, $\chi^2 = \sum_{i=1}^{6} (m_i/\delta
m_i)^2$\@.  The IDL Levenberg-Marquardt minimizer MPFIT \citep{mark09}
was employed to minimize $\chi^2$ in successive iterations.

\begin{figure*}[t!]
  \includegraphics[width=\linewidth]{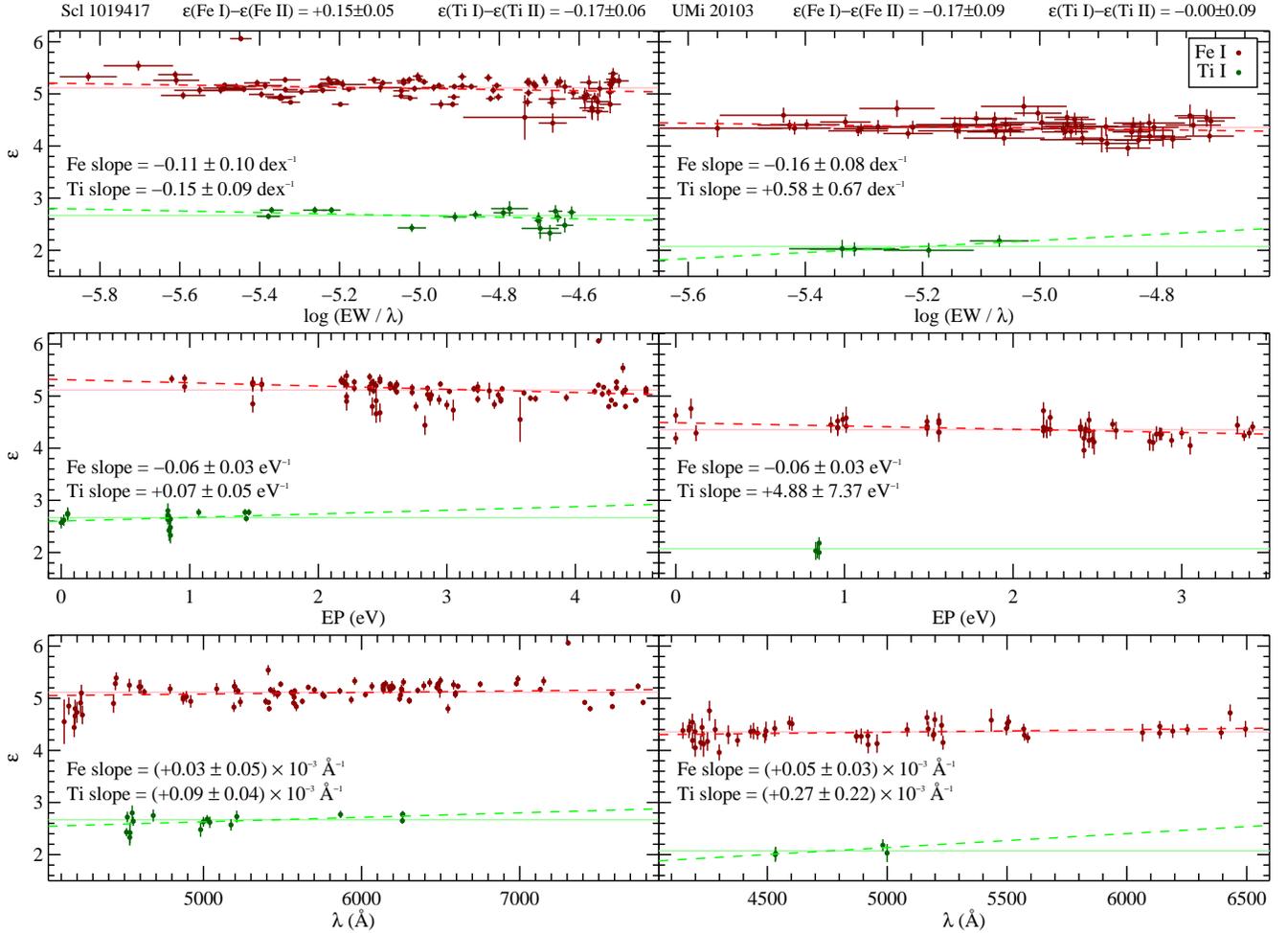}
  \caption{Abundances in Scl~1019417 and UMi~20103 from individual
    lines of \ion{Fe}{1} and \ion{Ti}{1} as a function of the lines'
    reduced width ($\rm{EW}/\lambda$, top), excitation potential
    (middle), and wavelength (bottom).  The error bars on the
    abundances are the standard deviations among 100 noise
    realizations of the spectrum.  The figure title gives the
    differences between the abundances of neutral and ionized species
    of Fe and Ti.  The best combination of effective temperature and
    microturbulent velocity minimizes the slopes of the relations in
    the top two panels and the differences in elemental abundances
    from different ionization states of Fe and Ti.  The confidence
    intervals on the slopes and differences between abundances of
    different ionization states were computed using a delete-1
    jackknife.\label{fig:atmpars}}
\end{figure*}

We performed this minimization using photometric and spectroscopic
temperatures.  For the adopted photometric temperature, the only
variable is $\xi$\@.  It was varied until $\chi^2$ was minimized.  For
the spectroscopic temperature, both \teff\ and $\xi$ were varied.
Table~\ref{tab:atmpars} gives the optimized parameters for both the
spectroscopic and photometric methods.  In order to estimate the
uncertainty introduced by spectral noise, we also computed the
atmospheric parameters for all 100 noise realizations of all four
spectra.  The standard deviations among 100 trials for each parameter,
$\delta_{\rm noise}$, are also listed in the table.

Figure~\ref{fig:atmpars} shows the trends of \ion{Fe}{1} and
\ion{Ti}{1} abundances with reduced width, excitation potential, and
wavelength for the spectroscopic temperatures in Scl~1019417 and
UMi~20103.  The lack of trends with reduced width and excitation
potential show that $\xi$ and \teff, respectively, have been measured
accurately.  The lack of a trend with wavelength is merely a check
that MOOG does not give different results as the continuous opacity
changes from the blue to red regions of the spectrum.

Although we experimented with measuring \logg\ from the spectrum, we
found that it was degenerate with \teff.  Particularly in the cases of
HD~115444 and Scl~1019417, the $\chi^2$ minimum was a long, narrow
valley in \teff--\logg\ space that permitted a temperature range of
hundreds of Kelvin.  On the other hand, \logg\ can be measured very
precisely from photometry in cases where the distance is known, such
as for the dSph stars.  Isochrones can pinpoint \logg\ to within 0.1.
Uncertainties in the photometry and distance modulus propagate to an
uncertainty in \logg\ of only 0.05.  Because Sculptor and Ursa Minor
have only ancient populations \citep[e.g.,][]{mon99,carr02},
uncertainty in the age of the star, say from 10 to 14~Gyr, contributes
negligible error to \logg\ (about 0.02).  Finally, the systematic
error in isochrone modeling may be quantified from the dispersion
among different isochrone sets.  The maximum difference between the
Yonsei-Yale, Victoria-Regina \citep{van06}, and Padova \citep{gir02}
isochrones is 0.12 for Scl~1019417 and 0.03 for UMi~20103.  Therefore,
we conclude that our abundance measurements are far more certain with
the photometric value of \logg\ than with a spectroscopic value.
Furthermore, we assert that the small uncertainty in \logg\ will
contribute insignificantly to uncertainty in the abundance
measurements.  Consequently, we do not consider \logg\ in our
abundance error estimates.

As a check on the surface gravity, the top of Figure~\ref{fig:atmpars}
gives the differences between \ion{Fe}{1} and \ion{Fe}{2} and between
\ion{Ti}{1} and \ion{Ti}{2}.  Ideally, these differences should be
zero.  However, non-local thermodynamic equilibirum (NLTE) effects can
alter the photospheric ionization balance.  \citet{the99} identified
overionization by ultraviolet radiation to be the most significant
NLTE effect for Fe abundance in metal-poor stars.  Essentially, the
true photosphere contains fewer \ion{Fe}{1} atoms (the minority
species) for its iron abundance than the idealized LTE photosphere.
The same effect applies to Ti, which has an ionization potential only
0.9~eV less than Fe.  Therefore, with a photometric surface gravity,
it may be impossible to find a combination of \teff\ and $\xi$ that
perfectly balances neutral and ionized species.  In fact, there may be
no choice of surface gravity that balances the abundances from
different ionization states.  The differences in
Figure~\ref{fig:atmpars} reflect this conundrum.  However, using a
photometric gravity insulates our abundance measurements from the
overionization effect.  For consistency with most of the literature,
we did not correct our \ion{Fe}{1} abundances for overionization.

The high precision in \logg\ afforded by photometry does not extend to
\teff.  Photometric uncertainties and systematic errors lead to
temperature errors exceeding 100~K\@.  The maximum difference between
the spectroscopic and photometric temperatures among the four stars in
our sample, shown in Table~\ref{tab:atmpars}, is 80~K\@.  The maximum
difference in $\xi$ between the two methods is just
0.03~km~s$^{-1}$\@.  We consider the spectroscopic \teff\ to be more
accurate than the photometric \teff.  For the remainder of this
article, abundances are derived using the atmospheric parameters
listed under ``Spectroscopic \teff'' in Table~\ref{tab:atmpars}.

\subsection{Abundance Measurements}
\label{sec:abundmeasure}

\addtocounter{table}{1}

\begin{figure}[t!]
  \includegraphics[width=\linewidth]{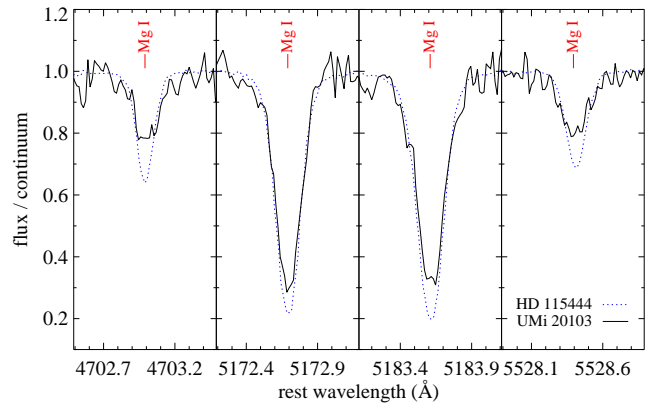}
  \caption{\ion{Mg}{1} lines in HD~115444 and UMi~20103.  These two
    stars have similar atmospheric parameters.  Therefore, the line
    strengths correspond fairly directly to the abundances.  We
    measured $\epsilon($\ion{Mg}{1}$) = 5.00$ and $4.81$ in HD~115444
    and UMi~20103, respectively.\label{fig:mg}}
\end{figure}

After the optimal atmospheric parameters were determined, we
interpolated the corresponding ATLAS9 model atmosphere from
\citeauthor{kir11a}'s (\citeyear{kir11a}) grid.  The abundances used
to compute opacities in the model atmosphere were not strictly
consistent with the abundances of the star.  We started with solar
abundances \citep[][except that $\epsilon(\rm{Fe}) = 7.52$]{and89}
scaled by \feh\@.  Then we rescaled the atmospheric abundances of O,
Ne, Mg, Si, S, Ar, Ca, and Ti by \afe, determined by the procedure
described in Section~\ref{sec:atm}.  This model atmosphere and the
line list were the inputs for MOOG\@.  Table~\ref{tab:abund} shows the
abundance results for the four stars.  The table also shows the
abundances relative to the solar abundances.\footnote{$\rm{[X/Y]} =
  (\epsilon(\rm{X})-\epsilon(\rm{Y})) -
  (\epsilon_{\sun}(\rm{X})-\epsilon_{\sun}(\rm{Y}))$}

Figure~\ref{fig:mg} gives an example of the abundance measurements.
It shows the four lines used to determine the \ion{Mg}{1} abundance in
UMi~20103.  It also shows the same lines in HD~115444, which has
$T_{\rm eff}$, $\log g$, and $\xi$ similar to UMi~20103.  Therefore,
line strengths correspond fairly directly to abundances.  The
\ion{Mg}{1} lines in UMi~20103 are slightly weaker than in HD~115444,
indicating that UMi~20103 has a slightly lower \ion{Mg}{1} abundance.
In fact, we measured $\epsilon($\ion{Mg}{1}$) = 5.00$ in HD~115444 and
$4.81$ in UMi~20103.

We also calculated abundances using the line lists from each of the
100 Monte Carlo noise realizations of the spectrum.  We estimated the
abundance uncertainty due to spectral noise for each line as the
standard deviation of the abundances from all 100 realizations.  We
call the abundance uncertainty on the $i^{\rm{th}}$ line
$\delta\epsilon_i$\@.  Then, we calculated the error-weighted mean
abundance from all $N$ lines of a given species:

\begin{equation}
\epsilon = \frac{\sum_{i=1}^{N} \epsilon_i \delta\epsilon_i^{-2}}{\sum_{i=1}^{N} \delta\epsilon_i^{-2}} \label{eq:epsilon}
\end{equation}

\noindent Each noise realization had its own value of $\epsilon$\@.
We estimated the uncertainty on $\epsilon$ due to spectral noise,
$\delta_{\rm{noise}}\epsilon$, as the standard deviation of all 100
values of $\epsilon$\@.

The error estimate $\delta_{\rm{noise}}\epsilon$ does not include
systematic error due to uncertainty in atomic parameters for each
line.  As a result, we also calculated the unbiased, weighted standard
deviation among the abundances from different lines of the same
species.  We divided this error estimate by the square root of the
number of lines, and we call it $\delta_{\rm{sys}}\epsilon$\@.  This
systematic error estimate came only from the unmodified spectrum and
not from any of the noise realizations.  The greater of
$\delta_{\rm{noise}}\epsilon$ and $\delta_{\rm{sys}}\epsilon$ is used
as the error bar in all figures.

Some elements and lines required special attention.  We detail them in
the following sections.

\subsection{Carbon}

\begin{figure}[t!]
  \includegraphics[width=\columnwidth]{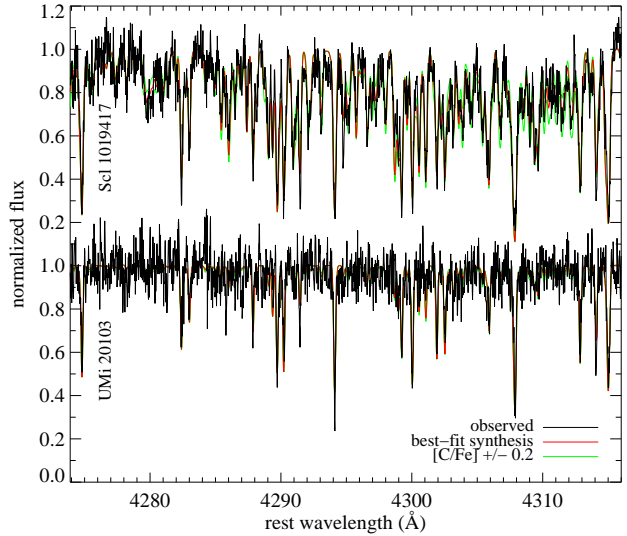}
  \caption{The CH region of the HIRES spectra of Scl~1019417 and
    UMi~20103 (thin black lines).  The best-fit synthetic spectra,
    when [C/H] was allowed to vary, are shown as thick red lines.
    Syntheses with [C/Fe] at 0.2~dex above and below the best-fit
    value are shown as green lines.\label{fig:ch}}
\end{figure}

We measured the abundance of neutral carbon from spectral synthesis of
the CH molecular G band.  First, we used MOOG to synthesize a spectrum
of the G band between 4273.9~\AA\ and 4333.0~\AA\@.  The line list came
from \citet{jor96}.  We assumed that the isotopic ratio
$^{12}\rm{C}/^{13}\rm{C} = 7.0$, but our carbon abundances changed
almost imperceptibly when we used different values.  Then, we computed
$\chi^2$ between the synthetic spectrum and the observed spectrum
divided by the DAOSPEC continuum.  The denominator of $\chi^2$ was the
flux error estimates from MAKEE.  Next, we adjusted the carbon
abundance until the $\chi^2$ was minimized.

We refined the continuum with the residuals of the fit, an approach
used for medium-resolution spectra by \citet{she09} and
\citeauthor*{kir10}.  We fit a B-spline with breakpoints every 500
pixels (10~\AA) to the quotient of the observed and best-fit synthetic
spectra.  Then, we divided the observed spectrum by this spline and
remeasured the best-fit carbon abundance.  We repeated this procedure
until the best-fit carbon abundance changed by less than 0.001~dex
between continuum iterations.

Figure~\ref{fig:ch} shows a region of the G band for Scl~1019417 and
UMi~20103.  The black lines are the observed spectra normalized by the
corrected continua.  The red lines are the best-fit synthetic spectra.
Despite the noise in the spectrum of UMi~20103, the formal error on
$\epsilon(\rm{C})$ is only 0.05~dex.  However, the G band is notorious
for systematic error in the line list and in \teff.  Therefore, we
assume a conservative systematic error of 0.20~dex.

\subsection{Oxygen}
\label{sec:oxygen}

The only oxygen line strong enough to be visible in the spectra of the
stars we observed is \ion{O}{1}~$\lambda$6300.3.  Unfortunately, the
radial velocities of Scl~1019417 and UMi~20103 are such that separate
telluric absorption lines fall exactly at 6300.3~\AA\ in the rest
frames.  The strengths of the telluric lines exceed the expected
strengths of the oxygen lines.  As a result, attempting to remove the
telluric absorption line would lead to a highly uncertain EW
measurement for the oxygen lines.  We estimated upper limits on oxygen
abundances from \ion{O}{1}~$\lambda$6364.

\subsection{Sodium}
\label{sec:sodium}

\begin{deluxetable}{lcccc}
\tablewidth{0pt}
\tablecolumns{5}
\tablecaption{NLTE Corrections\label{tab:nlte}}
\tablehead{\colhead{Species} & \colhead{HD~115444} & \colhead{HD~122563} & \colhead{Scl~1019417} & \colhead{UMi20103}}
\startdata
\ion{Na}{1} & $-0.39$ & $-0.32$ & $-0.06$ & $-0.20$ \\
\ion{Al}{1} & $+0.47$ & $+0.36$ & $+0.15$ & $+0.55$ \\
\ion{K}{1}  & $-0.36$ & \nodata & $-0.41$ & $-0.36$ \\
\enddata
\end{deluxetable}

\ion{Na}{1} was measured as described in
Section~\ref{sec:abundmeasure} from the Na~D lines.  However, we
corrected the abundance for NLTE effects.  \citet{lin11} calculated
such corrections for a large range of \teff, \logg, $\xi$, and line
strengths.  We used the grid and interpolation routine that K.~Lind
kindly provided to us to calculate NLTE corrections for our own
\ion{Na}{1} measurements as well as those of \citet{wes00}, who
measured \ion{Na}{1} from the doublet at 8190~\AA\@.
Table~\ref{tab:nlte} lists NLTE abundance corrections for our stars.

The minimum surface gravity in \citeauthor{lin11}'s (\citeyear{lin11})
correction table is $\mathlogg = 1.0$, larger than for Scl~1019417.
However, their Figure~4 shows that the Na~D NLTE correction is nearly
independent of \logg\ at 4300~K\@.  Therefore, we used the correction
for $\mathlogg = 1.0$.

\subsection{Aluminum}

We measured \ion{Al}{1} from the resonance line at 3962~\AA\@.  We did
not use \ion{Al}{1}~$\lambda$3944 because it is blended with CH lines.
The Al resonance lines are especially subject to NLTE corrections.
\citet{and08} computed such corrections for metal-poor stars.  We
consulted their Figure~2 to determine the appropriate correction for
each of our stars.  We linearly interpolated or extrapolated their
gravity- and metallicity-dependent corrections to estimate the
corrections appropriate for the atmospheric parameters of our stars.
We added this correction to the \ion{Al}{1} abundances.

\subsection{Potassium}

The strongest potassium line in the visible spectrum is
\ion{K}{1}~$\lambda$7699.  This is a strong resonance line highly
subject to NLTE effects.  \citet{iva00} modeled the potassium atom and
computed NLTE corrections to the \ion{K}{1} abundance.  We applied
corrections to all of our \ion{K}{1} abundances based on their
Figure~6.

\subsection{Hyperfine Splitting}
\label{sec:hfs}

Many of the elements in our repertoire (\ion{Sc}{2}, \ion{V}{1},
\ion{Mn}{1}, \ion{Mn}{2}, \ion{Co}{1}, \ion{Cu}{1}, \ion{Sr}{2},
\ion{Ba}{2}, \ion{La}{2}, \ion{Eu}{2}, and \ion{Pb}{1}) are subject to
hyperfine splitting of their energy levels.  None of the hyperfine
structure is so extended that the single-line fits are inappropriate.
Therefore, these absorption lines are best treated as blends of
multiple electronic transitions.  \citet{coh04} provided the atomic
data for the components of each blend.  We used the ``blends'' driver
of MOOG to compute the abundances.

\subsection{Upper Limits}

For species with only upper limits on EWs, we measured abundances from
all of the available upper limits.  For species with more than one
upper limit, the lowest abundance was used.

\subsection{Comparison to \protect \citet{wes00}}

\begin{figure*}[t!]
%\begin{figure}[t!]
  \includegraphics[width=\textwidth]{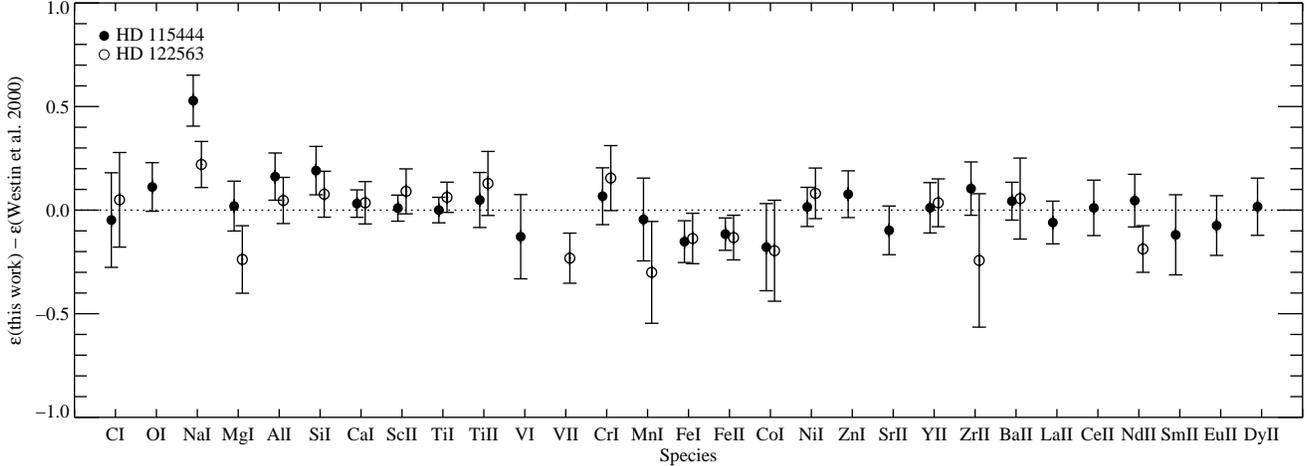}
  \caption{Differences between our and \protect\citeauthor{wes00}'s
    (\protect\citeyear{wes00}) abundance measurements for the two
    metal-poor abundance standards HD~115444 and HD~122563.  All
    species except Fe are given as relative to the abundance of Fe
    measured in the same ionization state.\label{fig:wes00_abund}}
%\end{figure}
\end{figure*}

Figure~\ref{fig:wes00_abund} shows the differences between our
measurements and those of \citet{wes00} for HD~115444 and HD~122563.
The error bars are the quoted errors from both studies added in
quadrature.  We corrected \citeauthor{wes00}'s \ion{Na}{1} abundance
measurement for NLTE effects as described in Section~\ref{sec:sodium}.

This comparison demonstrates that the new components of our technique
(DAOSPEC, a new version of MOOG, and consistent \afe\ between the
model atmosphere and the measured abundances) did not cause major
discrepancies with previous work.  Our measurements of \ion{Fe}{1} and
\ion{Fe}{2} are $\sim -0.15$~dex below those of
\citeauthor{wes00}\ Differences in \teff\ between the two studies do
not account for the difference because our measurements of \teff\ are
higher, which would lead to {\it larger} abundances.  Instead, the
difference likely comes from the different model atmosphere codes: our
use of ATLAS9 ``newodf'' versus \citeauthor{wes00}'s use of MARCS
\citep{gus75,edv93}.  We confirmed that the choice of model atmosphere
code is responsible for the shift in abudances by computing abundances
using the LTE version of MOOG (without an updated treatment of
scattering) with \citeauthor{wes00}'s line list, EWs, and atmospheric
parameters.  However, we used ATLAS9 model atmospheres instead of
MARCS.  The \ion{Fe}{1} abundances for both stars were indeed $\sim
0.15$~dex lower than \citeauthor{wes00}\ published.

The most discrepant species is \ion{Na}{1} (HD~115444, $\Delta\epsilon
= +0.5$).  We used the strong Na~D lines, whereas
\citeauthor{wes00}\ used the weaker \ion{Na}{1}~8190 doublet.
Interstellar absorption may contaminate the Na~D lines, but that
problem will not affect the dSph stars, which have larger absolute
radial velocities.

\subsection{Comparison to DEIMOS Measurements}

\begin{deluxetable*}{lcccccccc}
%\begin{deluxetable}{lcccccccc}
%\rotate
\tabletypesize{\small}
\tablewidth{0pt}
\tablecolumns{9}
\tablecaption{Comparison of DEIMOS and HIRES Abundance Measurements\label{tab:deimos}}
\tablehead{\colhead{Spectrograph} & \colhead{\teff~(K)} & \colhead{\logg} & \colhead{$\xi$~(km~s$^{-1}$)} & \colhead{[\ion{Fe}{1}/H]} & \colhead{[\ion{Mg}{1}/\ion{Fe}{1}]} & \colhead{[\ion{Si}{1}/\ion{Fe}{1}]} & \colhead{[\ion{Ca}{1}/\ion{Fe}{1}]} & \colhead{[\ion{Ti}{1}/\ion{Fe}{1}]}}
\startdata
\cutinhead{Scl 1019417}
DEIMOS & $4175 \pm 61$ & $0.49 \pm 0.08$ &       2.03      & $-2.46 \pm 0.11$ & $+0.69 \pm 0.16$ & $+0.62 \pm 0.18$ & $+0.54 \pm 0.16$ & $+0.20 \pm 0.13$ \\
HIRES  & $4280 \pm 32$ & $0.42 \pm 0.08$ & $2.21 \pm 0.10$ & $-2.40 \pm 0.05$ & $+0.57 \pm 0.07$ & $+0.53 \pm 0.04$ & $+0.19 \pm 0.04$ & $+0.08 \pm 0.07$ \\
\cutinhead{UMi 20103}
DEIMOS & $4824 \pm 86$ & $1.66 \pm 0.05$ &       1.75      & $-3.62 \pm 0.35$ &     \nodata      &     \nodata      & $+0.89 \pm 0.48$ & $+0.69 \pm 0.84$ \\
HIRES  & $4799 \pm 72$ & $1.66 \pm 0.05$ & $2.04 \pm 0.27$ & $-3.16 \pm 0.10$ & $+0.39 \pm 0.09$ & $+0.38 \pm 0.21$ & $+0.22 \pm 0.08$ & $+0.25 \pm 0.10$ \\
\enddata
\tablecomments{DEIMOS measurements are taken from \protect \citeauthor*{kir10}.  The errors on the HIRES atmospheric parameters are the random noise error terms only (Section~\protect\ref{sec:atm}).  The errors on the HIRES abundances are the larger of $\delta_{\rm noise}$ or $\delta_{\rm sys}$ (Section~\protect\ref{sec:abundmeasure}).}
%\end{deluxetable}
\end{deluxetable*}

Table~\ref{tab:deimos} shows the DEIMOS medium-resolution
(\citeauthor*{kir10}) and HIRES high-resolution measurements for the
two dSph stars.  The HIRES measurements roughly agree with the DEIMOS
measurements.  The differences in \feh\ for Scl~1019417 and UMi~20103
are 0.06~dex ($0.5\sigma$) and 0.46~dex ($1.3\sigma$), respectively.
Both differences are in the sense that the HIRES measurements are
larger.  These two stars alone do not indicate that the DEIMOS
measurements tend to be too metal-poor.  \citeauthor*{kir10} compared
DEIMOS abundances to the high-resolution literature for 132 stars.
They found no systematic offset nor any trend of $\Delta\mathfeh$ with
\feh\@.  The differences between HIRES and DEIMOS here are consistent
with random noise.

It was not possible to measure with useful precision any elements
other than Fe in the DEIMOS spectrum of UMi~20103.  It was possible to
measure [Mg/Fe], [Si/Fe], [Ca/Fe], and [Ti/Fe] in the DEIMOS spectrum
of Scl~1019417.  The differences are $-0.12$~dex ($0.7\sigma$),
$-0.09$~dex ($0.5\sigma$), $-0.35$~dex ($2.1\sigma$), and $-0.12$~dex
($0.8\sigma$), all in the sense that the HIRES measurements are lower.
Again, it is impossible to draw conclusions from a single-star
comparison.  We defer to \citeauthor*{kir10}'s comparisons for a
complete view of the accuracy of the DEIMOS measurements.

%%%%%%%%%%%%%%%%%%%%%%%%%%%%%%%%%
%%%%%%%%%   SECTION 5   %%%%%%%%%
%%%%%%%%%%%%%%%%%%%%%%%%%%%%%%%%%

\section{Discussion}
\label{sec:discussion}

\begin{figure*}[t!]
  \centering
  \includegraphics[width=0.95\linewidth]{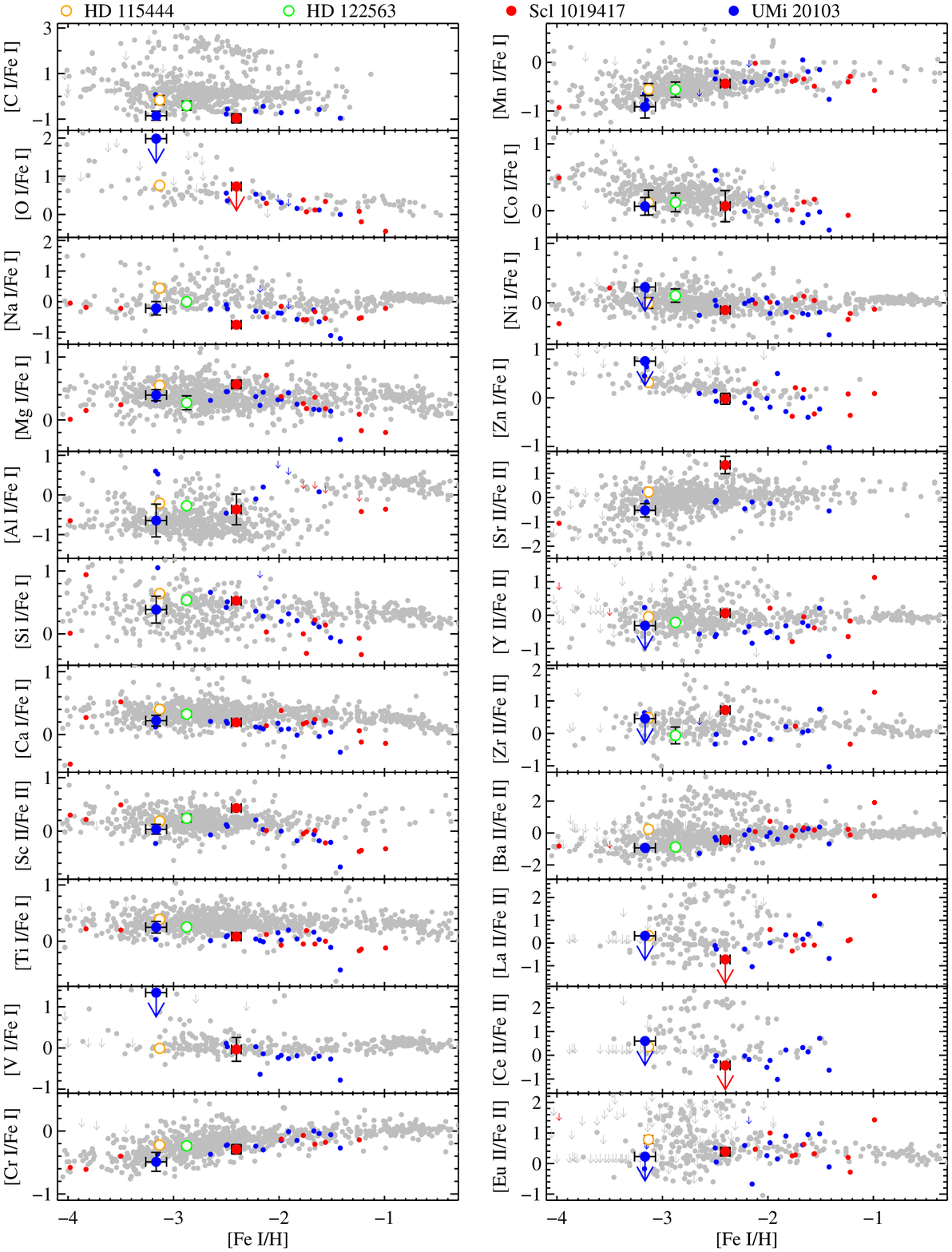}
  \caption{Abundances in dSph stars compared to the MW halo.  The
    program stars are shown as large, solid, colored points, and the
    halo abundance standards are shown as large, hollow, colored
    points.  For comparison, other halo stars (gray points) are shown
    from \protect \citeauthor{frebel10}'s (\protect
    \citeyear{frebel10}) compilation.  Previous studies of Sculptor
    \protect \citep{she03,gei05,fre10a,taf10} and Ursa~Minor \protect
    \citep{she01,sad04,coh10} are shown as small red and blue points,
    respectively.  Error bars represent the larger of the noise or
    systematic error.  Downward pointing arrows represent $2\sigma$
    upper limits.  NLTE corrections have been applied to our stars
    (but not necessarily the comparison sample from the literature)
    for \ion{Na}{1} and \ion{Al}{1}.\label{fig:abund}}
\end{figure*}

Figure~\ref{fig:abund} shows the abundances of the two program stars,
Scl~1019417 and UMi~20103, along with the halo abundance standards,
HD~115444 and HD~122563.  For context, the figure also shows previous
literature measurements for Sculptor, Ursa Minor, and the MW halo.  In
this section, we offer some commentary on the particular abundances of
our two program stars.

\subsection{The DSphs in Context}

The abundance ratios we measured are not especially peculiar for very
metal-poor dSph stars.  These abundance ratios have been discussed at
length in the literature.  \citet{tol09} and \citet{mcw10} have
written recent reviews on nucleosynthesis, in particular as it
concerns local galaxies.

Examining the detailed abundance patterns at a range of metallicities
affords a fuller appreciation of the chemical evolution of a dSph.  To
first order, a sequence in metallicity is a sequence in time.  As time
advances, more supernovae (SNe) explode to enrich the metallicity of
the dSph's interstellar medium.  Therefore, the later-forming stars
have higher metallicities.  This process may be modeled quantitatively
and in detail \citep[e.g.,][]{iku02,lan04,kir11b}.

We compiled high-resolution abundance measurements for Sculptor and
Ursa Minor from the published literature.  Where necessary, we shifted
their abundances to match our solar abundance scale. These are shown
in Figure~\ref{fig:abund} to provide context for our own measurements.
\citet{she03} first observed Sculptor at high-resolution.  They
obtained VLT/UVES spectroscopy of five red giants at $R = 40\,000$
with $\rm{SNR} \sim 30$~pixel$^{-1}$\@.  \citet{gei05} later obtained
UVES spectra for four additional red giants at $R = 20\,000$ and
$\rm{SNR} \sim 90$~pixel$^{-1}$\@.  \citet{fre10a} observed the
extremely metal-poor star Scl~1020549, with Magellan-Clay/MIKE at $R =
33\,000$ and $\rm{SNR} \sim 35$~pixel$^{-1}$\@.  \citet{taf10}
observed two additional extremely metal-poor stars in Sculptor,
including the most metal-poor star known in any dSph, with VLT/UVES at
$R = 40\,000$ with $\rm{SNR} \sim 35$~pixel$^{-1}$\@.  \citet{she01}
obtained the first high-resolution spectra in Ursa Minor.  They
observed six red giants with Keck/HIRES at $R = 34\,000$ and $\rm{SNR}
\sim 30$~pixel$^{-1}$\@.  One of these, Ursa Minor~K, is a carbon
star, and we do not plot it.  \citet{sad04} observed three red giants
with Subaru/HDS at $R = 45\,000$ and $\rm{SNR} \sim
55$~pixel$^{-1}$\@.  Two of these three stars overlapped
\citeauthor{she03}'s (\citeyear{she03}) sample.  Given
\citeauthor{sad04}'s higher resolution and SNR, we adopted their
abundances instead of \citeauthor{she03}'s abundances.  Finally
\citet{coh10} observed 10 red giants in Ursa Minor with HIRES at $R =
35\,000$ and $\rm{SNR} \sim 90$ per resolution element.  None of these
stars overlapped with the other samples, and they are all shown in
Figure~\ref{fig:abund}.

As has been noted many times previously, \afe\ ([O/Fe], [Mg/Fe],
[Si/Fe], and [Ca/Fe]) in the dSphs declines with increasing \feh\ at
$\mathfeh \ga -2.5$\@.  The decrease arises from a growing contribution
of the products of Type~Ia SNe compared to Type~II (core collapse)
SNe.  The former produce iron peak elements and virtually no $\alpha$
elements, whereas the latter produce $\alpha$ elements and somewhat
less iron.  The timescale for Type~II SNe is 3--20~Myr after star
formation, and the minimum delay for Type~Ia SNe is about 60~Myr
\citep[e.g.,][]{tho11}.  Therefore, a galaxy begins its life with the
low \feh\ and high \afe\ imposed by Type~II SNe, but gradually,
Type~Ia SNe depress this ratio as \feh\ increases due to the explosion
of both types of SNe.  This simplistic description assumes that the
galaxy does not experience any sudden star formation bursts that can
produce many Type~II SNe, causing an uptick in
\afe\ \citep[see][]{gil91}.

If Type~Ia SNe experience a non-zero delay time longer than the delay
for Type~II SNe, then dSphs should exhibit a plateau of high \afe\ at
low \feh, corresponding to stars formed before the advent of Type~Ia
SNe.  \citet{coh10} identified such a plateau in Ursa Minor.  Although
the sparse sampling led to different maximum values of \feh\ for the
plateau depending on the particular $\alpha$ element, Type~Ia SNe seem
to have begun to affect Ursa Minor's abundance pattern somewhere in
the range $-3.0 \le \mathfeh \le -2.5$\@.  This is consistent with the
conclusion of \citet{kir11b}, who found that no dSph has a high
\afe\ plateau that extends to $\mathfeh > -2.5$, except for [Ca/Fe] in
Sculptor.

The existence of an \afe\ plateau in Sculptor is controversial.  The
\afe\ ratios measured by \citet{she03} and \citet{gei05} decline
monotonically with metallicity.  Measurements from the Dwarf
Abundances and Radial Velocities Team show a plateau in [Ca/Fe] ending
at $\mathfeh = -1.8$ \citep{tol09}.  \citet{kir09,kir11b} confirmed
this result with medium-resolution spectroscopy of 376 Sculptor red
giants.  However, \citeauthor{kir11b}\ also measured [Mg/Fe] and
[Si/Fe] and found no evidence of a plateau in those $\alpha$ elements.
\citet{tol09} also presented [Mg/Fe] measurements in Sculptor.  The
dispersion in [Mg/Fe] at fixed [Fe/H] increases below $\mathfeh =
-1.8$, but evidence for a plateau in [Mg/Fe] is inconclusive.  Our own
measurements of Scl~1019417 show that [Mg/Fe] and [Si/Fe] are
consistent with rising toward lower metallicity, whereas [Ca/Fe]
remains in the previously established plateau.  Therefore, the
distribution of [Ca/Fe] in Sculptor seems more complex than for
[Mg/Fe] or [Si/Fe].

The most metal-poor stars in Sculptor (Scl~1020549,
\citeauthor{fre10a}\ \citeyear{fre10a}; Scl07-49 and Scl07-50,
\citeauthor{taf10}\ \citeyear{taf10}) do not conform to this simple
picture.  Their [Mg/Fe] ratios are lower than the typical [Mg/Fe]
ratio at ${\rm [Fe/H]} \sim -2$, and their [Si/Fe] and [Ca/Fe] span a
large range.  These stars are so metal-poor ($\mathfeh \la -3.5$) that
they could be the direct products of one or a few Population~III SNe.
If so, then their abundances would reflect the specific mass- and
explosion energy-dependent yields of those particular SNe.  It is only
when the SNe yields are averaged over progenitor mass and explosion
energies that the global \afe\ interpretation as the balance of
Types~II and Ia SNe makes sense.

\begin{figure}[t!]
  \centering
  \includegraphics[width=0.95\linewidth]{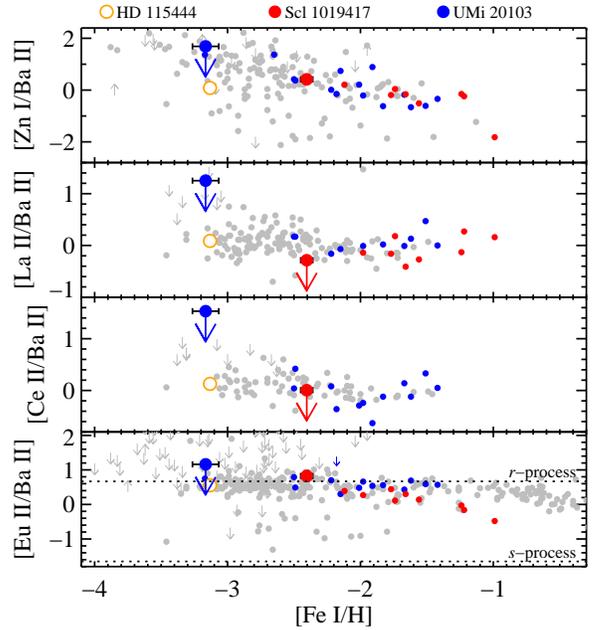}
  \caption{Neutron-capture abundance ratios.  The solar system [Eu/Ba]
    ratios from the $s$- and $r$-processes \citep{sim04} are shown as
    dotted lines in the bottom panel.  The symbols have the same
    meanings as in Figure~\ref{fig:abund}.\label{fig:ncap}}
\end{figure}

The neutron-capture elements also trend with \feh\@.  Roughly
speaking, the $r$-process dominates the production of neutron-capture
elements at early times (low \feh).  After hundreds of Myr, thermally
pulsating asymptotic giant branch (AGB) stars can expel $s$-process
material.  The ratio [Eu/Ba] is a diagnostic of the relative
contributions of the $r$- and $s$-processes.  Because Eu is produced
mostly in the $r$-process, [Eu/Ba] decreases as more $s$-process
sources contribute to the Ba content of the star.  This phase is
especially apparent in Fornax, where [Eu/Ba] decreases steeply as
[Fe/H] increases \citep{let10}.  The decrease of [Eu/Ba] in Sculptor
continues the trend established by Fornax, but at lower [Fe/H]
\citep{tol09}.  The bottom panel of Figure~\ref{fig:ncap} shows the
abundances of [Eu/Ba] in Sculptor, Ursa Minor, and the MW halo.  The
$s$-process did not contribute to MW stars by lowering [Eu/Ba] until
${\rm [Fe/H]} > -1$, but it began in Sculptor at ${\rm [Fe/H]} \sim
-1.8$\@.  Scl~1019417 is the lowest metallicity star in Sculptor with
measurements of both Ba and Eu, and it shows that no $s$-process
material was present in Sculptor when the dSph's metallicity was
${\rm[Fe/H]} = -2.5$\@.  The differences between the MW and Sculptor
suggest that the MW progenitors reached a higher metallicity sooner
than Sculptor.  In contrast, the [Eu/Ba] in Ursa Minor does not
decrease with [Fe/H]\@.  Of course, AGB stars in Ursa Minor did
produce $s$-process nuclei, just like in any other dSph.  However, the
star formation in Ursa Minor did not last long enough for the
$s$-process material to be incorporated into stars.  Alternatively,
the star formation rate (SFR) was so low by the time AGB stars began
to spew $s$-process nuclei that the chances of finding such a star are
low.

Star formation histories derived by chemical evolution models support
these interpretations.  For Sculptor, the SFR began to decline 300~Myr
after the first star was born in Sculptor \citep{kir11b}, but some
star formation proceeded for up to 7~Gyr \citep{deb12}.  This would
allow some AGB products to have polluted the more metal-rich stars in
Sculptor while leaving the metal-poor stars free of the $s$-process.
The star formation duration in Ursa Minor was as short as 400~Myr
\citep{kir11b}, which would ensure that virtually no stars were
contaminated by AGB ejecta.

\subsection{Comparison of DSphs to the Milky Way Halo}

Metal-poor stars in dSphs provide constraints on dSphs' contribution
to the MW halo.  Even if the surviving dSphs are not representative of
the dominant inner halo constituents \citep[early-accreted
  dIrrs,][]{rob05}, the most metal-poor stars would presumably have
abundance patterns consistent across all types of dwarf galaxy.  The
earliest SNe do not have foreknowledge of the dwarf galaxy's final
stellar mass or its fate (accretion or survival).  Therefore, the
earliest forming, most metal-poor stars should have similar abundance
patterns.

In order to compare Sculptor's and Ursa Minor's abundance patterns to
the halo, we show \citeauthor{frebel10}'s (\citeyear{frebel10})
compilation\footnote{Figures~\ref{fig:abund} includes halo star
  abundance measurements from the following sources, compiled by
  \citet{frebel10}:
  \citet{aok00,aok01,aok02a,aok02b,aok02c,aok02d,aok05,aok06,aok07a,aok07b,aok07c,aok08,arn05,bar05b,barb05,bon09,bur00,car02,cay04,chr04,coh03,coh04,coh06,coh07,coh08,cow02,fra07,fre07a,fre07b,ful00,hay09,hil02,hon04,hon06,hon07,ito09,iva03,iva05,iva06,joh02,joh01,joh02a,joh02b,joh04,jon05,jon06,lai07,lai08,lai09,luc03,mas06,mcw98,mcw95,nor97a,nor97b,nor97c,nor00,nor01,nor02,nor07,ple04,pre00,pre01,pre06,roe08,rya91,rya96,siv04,siv06,sne03,spi00,spi05,wes00};
  and \citet{zac98}.  DSph stars and stars with $\mathfeh < -4$ in
  \citeauthor{frebel10}'s compilation are not included.} of halo star
abundances as gray points in Figures~\ref{fig:abund} and
\ref{fig:ncap}.  The abundances have been shifted to match our solar
abundance scale.  Note that high-resolution spectroscopy imposes a
heavy bias toward bright, nearby stars.  Therefore, the large majority
of stars in this halo sample belong to the inner halo.

The dSph stars show the familiar discrepancy with the halo stars at
$\mathfeh \ga -2$, particularly for the $\alpha$ elements
\citep{tol03,ven04}.  The more metal-poor dSph stars, including the
two stars that comprise our sample, have abundance ratios broadly
consistent with the inner halo, as previously noted by
\citet{fre10a,fre10b}, \citet{sim10}, \citet{nor10}, and
\citet{taf10}.

However, some elements show deviation worth mentioning.  First, carbon
in both Scl~1019417 and UMi~20103 ($\rm{[C/Fe]} \approx -0.9$) is
lower than the locus defined by halo stars.  This is merely a
consequence of the dSph sample, which consists exclusively of evolved
red giants.  Red giants dredge up carbon-depleted material as they
evolve up the red giant branch \citep{sun81}.

However, the dredge-up mechanism does not explain the low abundances
of Na and some neutron-capture elements in both of our stars.
Dredge-up does not affect Na abundances in metal-poor stars nearly as
much as it affects C abundances and isotopic ratios
\citep[e.g.,][]{spi06}.  Even if dredge-up were to affect the Na
abundances in metal-poor giants, it would increase [Na/Fe].  On the
other hand, [Na/Fe] in metal-poor stars in dSphs is low compared to
most halo stars.  In fact, Scl~1019417 has one of the lowest [Na/Fe]
ratios of any metal-poor star.  Some neutron-capture elements in dSphs
also show mild deviation from the halo.  At ${\rm [Fe/H]} < -2$,
[Zn/Fe], [Ba/Fe], [La/Fe], and [Ce/Fe] tend to lie along the lower
edge of the envelope defined by halo stars, as shown in
Figure~\ref{fig:abund}.  Low [Ba/Fe] ratios in extremely metal-poor
stars in classical and ultra-faint dSphs have also been noted by
\citet{she01}, \citet{ful04}, \citet{koc08}, \citet{tol09}, and
\citet{fre10b}.  The ratios of these neutron-capture elements to each
other (Figure~\ref{fig:ncap}) do not appear particularly discrepant,
except perhaps for the low upper limit on [La/Ba] in Scl~1019417.

We suggest that the origin for these deviations is the metallicity
dependence of these elements' SN yields.  The production of Na in
Type~II SNe depends sensitively on the neutron excess, which depends
on the initial metallicity of the exploding star \citep{woo95}.  Lower
metallicity SNe produce lower [Na/Fe] ratios.  The metallicity
dependence of SN yields can explain the distribution of light elements
in the MW bulge \citep{tsu02b,lec07}.  The metallicity dependence of
the yields of neutron-capture elements is less clear.  However, the
$r$-process seems to consist of at least two components: the main
$r$-process, which produces the full range of atomic number, and the
weak $r$-process, which produces elements lighter than Ba
\citep{ish05}.  (Also see \citeauthor{tra04}\ \citeyear{tra04} and
\citeauthor{qia07}\ \citeyear{qia07}.)  \citet[][their
  Section~5.8.2]{taf10} summarized the relation of the multiple
components of the $r$-process to the observed abundances of extremely
metal-poor stars in dwarf galaxies.  Although the metallicity
dependence of the yields of $r$-process elements is not
well-understood, at least one potential source of the
$r$-process---the lighter element primary process \citep{tra04}---is
expected to occur only in low-metallicity stars.  Therefore, it is
conceivable that the $r$-process yields have some metallicity
dependence.

The inner halo progenitors were presumably massive galaxies
\citep{rob05} that began their lives with a great deal of gas.  As a
result, the first SNe did not enrich the galaxy very much.
Consequently, the abundance patterns seen at, say, $\mathfeh = -2.5$
are reflective of SNe with metallicities only slightly less than
$\mathfeh = -2.5$\@.  On the other hand, the surviving dSphs are small
and dark matter-dominated.  The gradual inflow of gas at early times
\citep{kir11b} ensures that the first SN will pollute a small amount
of gas.  As a result, a single SN can enrich the entire protogalaxy to
as much as $Z = 10^{-3}~Z_{\sun}$ \citep{wis12}.  Therefore, the gap
between a SN's metallicity and the metallicity of the stars that
formed from its ejecta is larger than in the massive inner halo
progenitors.  The abundance pattern of a star at $\mathfeh = -2.5$ in
a small dSph could reflect the yields of a SN with $\mathfeh = -3$ or
even lower.  Therefore, the metallicity-dependent rise of Na and the
$r$-process elements is deferred to stars with higher \feh\ in the
dSphs than in the more massive halo progenitors.

Our suggestion is related to \citeauthor{mcw03}'s (\citeyear{mcw03})
conclusion that metallicity-dependent yields caused the evolution of
[Mn/Fe] in the Sagittarius dSph to differ from the MW bulge.  Whereas
our suggestion involves the metallicity dependence of the Type~II SN
yields of Na and possibly the neutron-capture elements at $\mathfeh <
-2$, \citeauthor{mcw03}'s explanation concerns the
metallicity-dependence of the Type~Ia SN yield of Mn at $\mathfeh \sim
-1$.  Also, our hypothesis explains the different abundance patterns
between the dSphs and the MW as a result of different masses of the
star-forming gas polluted by SNe.  In contrast, \citet{mcw03}
theorized that the difference is a result of the slower chemical
evolution of Sagittarius compared to the MW bulge.

We emphasize that the abundance patterns of very metal-poor stars in
dSphs do not lie outside of the boundary defined by halo stars.
Rather, some abundance ratios lie on or near the boundary.  Therefore,
some very metal-poor stars in the high-resolution halo samples, which
are heavily biased toward the inner halo, likely came from galaxies
very similar to the surviving dSphs.  Other very metal-poor halo stars
with higher [Na/Fe] and [$n$/Fe] ratios came from different types of
galaxies.  If our suggestion concerning the metallicity-dependence of
massive SN yields is correct, then these stars came from galaxies more
massive that the surviving dSphs.  Alternatives to
metallicity-dependent yields include inhomogeneous mixing
\citep{oey00,tsu02a} and mass-dependent yields coupled with stochastic
sampling of the initial mass function (D.~Lee et al., in preparation).

%%%%%%%%%%%%%%%%%%%%%%%%%%%%%%%%%
%%%%%%%%%   SECTION 6   %%%%%%%%%
%%%%%%%%%%%%%%%%%%%%%%%%%%%%%%%%%

\section{Summary and Conclusions}
\label{sec:conclusions}

From Keck/DEIMOS medium-resolution spectroscopy, \citet{kir10}
discovered several very metal-poor red giants in the MW dSphs. We
observed two of these (one in Sculptor and one in Ursa Minor) at
high-resolution with Keck/HIRES\@.  We measured the detailed
abundances of these stars.  Because one of them is faint ($V = 18.4$),
we paid careful attention to the abundance uncertainties introduced by
Poisson noise in the spectrum.  We measured the abundances from 100
different noise realizations of the spectra.  This approach required
automated EW measurements.  In order to check the accuracy of our
results, we applied the same technique to two well-studied halo
abundance standards.  We found that our measurements were consistent
with those of \citet{wes00}.

Our high-resolution measurements roughly confirm the DEIMOS values for
\feh, although the HIRES measurement of \feh\ for UMi~20103 is
0.46~dex ($1.3\sigma$) higher than the DEIMOS value.  The HIRES
metallicites are [\ion{Fe}{1}$\rm{/H]} = -3.16$ for UMi~20103 and
[\ion{Fe}{1}$\rm{/H]} = -2.40$ for Scl~1019417.  We also confirmed the
DEIMOS measurements of [$\alpha$/Fe] ratios in Scl~1019417, although
the HIRES [$\alpha$/Fe] ratios are slightly lower than the DEIMOS
measurements.

The abundance patterns of these two stars support previously
established trends.  In particular, the elemental abundance pattern of
very metal-poor dSph stars is largely consistent with very metal-poor
MW halo stars.  However, certain element ratios ([Na/Fe], [Zn/Fe],
[Ba/Fe], [La/Fe], and [Ce/Fe]) in Sculptor and Ursa Minor lie only at
the lower envelope of abundance ratios defined by inner halo stars.
We suggest that the earliest SNe in dSphs pollute a smaller gas mass
than the earliest SNe in the inner halo progenitors.  As a result, the
metallicity dependence of SN yields defers the rise in these ratios to
higher metallicities in the dSphs than in the inner halo.

This explanation affirms that objects identical to the surviving dSphs
were not the building blocks of the inner halo, which formed roughly
10~Gyr ago.  Instead, the surviving dSphs and objects like them are
actively building the outer halo.  \citet{roe09} showed that the outer
halo is chemically distinct from the inner halo.  Therefore, the two
entities must have had different formation mechanisms, even at the
lowest metallicities.  \citet{fon06a,fon06b} modeled the distribution
of \feh\ and \afe\ in the different spatial and kinematic components
of a MW-like halo.  A fair test of the hierarchical model would be to
compare their predicted abundance patterns to observations of the
outer halo, not to compare surviving dSphs to the inner halo.  We
anticipate that future missions such as GAIA \citep{jor08} will settle
this issue permanently.

\acknowledgments We thank the referee for a careful report that
greatly improved the quality of our study.  We also thank K. Lind for
providing a customized table of NLTE correction for sodium.  Support
for this work was provided by NASA through Hubble Fellowship grant
51256.01 awarded to ENK by the Space Telescope Science Institute,
which is operated by the Association of Universities for Research in
Astronomy, Inc., for NASA, under contract NAS 5-26555.  J.G.C. thanks
NSF grant AST-0908139 for partial support.

We are grateful to the many people who have worked to make the Keck
Telescope and its instruments a reality and to operate and maintain
the Keck Observatory.  The authors wish to extend special thanks to
those of Hawaiian ancestry on whose sacred mountain we are privileged
to be guests.  Without their generous hospitality, none of the
observations presented herein would have been possible.

{\it Facility:} \facility{Keck:I (HIRES)}

\clearpage
\tabletypesize{\scriptsize}
\begin{turnpage}

\renewcommand{\thetable}{\arabic{table}}
\setcounter{table}{0}
\begin{deluxetable}{lccccccccccccccc}
\tabletypesize{\scriptsize}
\tablewidth{0pt}
\tablecolumns{15}
\tablecaption{HIRES Observations and Data Quality\label{tab:obs}}
\tablehead{\colhead{Star} & \colhead{RA (J2000)} & \colhead{Dec (J2000)} & \colhead{$V$} & \colhead{$V-I$} & \colhead{$V-J$} & \colhead{$V-K$} & \colhead{UT Date} & \colhead{Exposures} & \colhead{Tot. Exp.} & \colhead{Seeing} & Range (\AA) & \colhead{$R$\tablenotemark{a}} & \colhead{SNR\tablenotemark{b}} & \colhead{$v_r$ (km~s$^{-1}$)\tablenotemark{c}}}
\startdata
\cutinhead{Program Stars}
Scl 1019417 & $01^{\mathrm{h}} 01^{\mathrm{m}} 42^{\mathrm{s}}$ & $-33\arcdeg 43\arcmin 09\arcsec$ & 16.93 &  1.39 & \nodata & \nodata & 2010 Nov 27    & $8 \times 30$~min & 240~min & $1 \farcs 4$ & 3927--8362 & $29\,300$ &     105 &  \phs $98.87 \pm 0.08$ \\
UMi 20103   & $15^{\mathrm{h}} 09^{\mathrm{m}} 58^{\mathrm{s}}$ & $+67\arcdeg 09\arcmin 28\arcsec$ & 18.41 &  1.05 & \nodata & \nodata & 2010 Apr 2\phn & $8 \times 30$ + $2 \times 24$~min           & 438~min & $0 \farcs 7$ & 3927--8362 & $34\,500$ & \phn 91 &     $-244.48 \pm 0.08$ \\
            &                                                   &                                  &       &       &         &         & 2011 Jun 6\phn & $2 \times 30$~min & & $1 \farcs 0$ & & & & \\
            &                                                   &                                  &       &       &         &         & 2012 Feb 2\phn & $3 \times 30$~min & & $0 \farcs 7$ & & & & \\
\cutinhead{Abundance Standards}
HD 115444   & $13^{\mathrm{h}} 16^{\mathrm{m}} 42^{\mathrm{s}}$ & $+36\arcdeg 22\arcmin 53\arcsec$ &  9.00 &  1.09 &  1.84   &  2.39   & 2005 Jun 16    & 450~s + 250~s & 700~s & $0 \farcs 7$ & 3871--8364 & $37\,300$ &     566 & \phn $-25.80 \pm 0.06$ \\
HD 122563   & $14^{\mathrm{h}} 02^{\mathrm{m}} 32^{\mathrm{s}}$ & $+09\arcdeg 41\arcmin 10\arcsec$ &  6.19 &  1.40 &  1.86   &  2.50   & 2006 Apr 16    & $2 \times 60$~s & 120~s    & $0 \farcs 6$ & 3185--5993 & $31\,100$ &     356 & \phn $-24.67 \pm 0.07$ \\
\enddata
\tablenotetext{a}{Resolving power, defined as the FWHM of unbroadened spectral features divided by wavelength.  This number depends on both the spectrograph configuration and stellar macroturbulence.}
\tablenotetext{b}{Signal-to-noise ratio per FWHM resolution element at 5750~\AA\@.}
\tablenotetext{c}{Heliocentric radial velocity.}
\end{deluxetable}

\begin{deluxetable}{lccrccccccccccccccc}
\tablecolumns{19}
\tablewidth{0pt}
\tablecaption{Line List with Equivalent Widths\label{tab:ew}}
\tablehead{ & & & & \multicolumn{3}{c}{HD 115444} & & \multicolumn{3}{c}{HD 122563} & & \multicolumn{3}{c}{Scl 1019417} & & \multicolumn{3}{c}{UMi 20103} \\ \cline{5-7} \cline{9-11} \cline{13-15} \cline{17-19}
\colhead{Species} & \colhead{Wavelength} & \colhead{EP} & \colhead{$\log gf$} & \colhead{EW} & \colhead{$\delta$EW$_{\rm noise}$} & \colhead{$\delta$EW$_{\rm DAO}$} & & \colhead{EW} & \colhead{$\delta$EW$_{\rm noise}$} & \colhead{$\delta$EW$_{\rm DAO}$} & & \colhead{EW} & \colhead{$\delta$EW$_{\rm noise}$} & \colhead{$\delta$EW$_{\rm DAO}$} & & \colhead{EW} & \colhead{$\delta$EW$_{\rm noise}$} & \colhead{$\delta$EW$_{\rm DAO}$} \\
& \colhead{(\AA)} & \colhead{(eV)} & & \colhead{(m\AA)} & \colhead{(m\AA)} & \colhead{(m\AA)} & & \colhead{(m\AA)} & \colhead{(m\AA)} & \colhead{(m\AA)} & & \colhead{(m\AA)} & \colhead{(m\AA)} & \colhead{(m\AA)} & & \colhead{(m\AA)} & \colhead{(m\AA)} & \colhead{(m\AA)}}
\startdata
\ion{Li}{1} & 6707.76 & 0.000 & $ -0.002$ &  \nodata & \nodata & \nodata & &  \nodata & \nodata & \nodata & & $<  4.8$ & \nodata & \nodata & & $< 12.4$ & \nodata & \nodata \\
\ion{O}{1}  & 6300.30 & 0.000 & $ -9.780$ & $   3.4$ &  $ 0.3$ & \nodata & &  \nodata & \nodata & \nodata & &  \nodata & \nodata & \nodata & &  \nodata & \nodata & \nodata \\
\ion{O}{1}  & 6363.78 & 0.020 & $-10.300$ &  \nodata & \nodata & \nodata & &  \nodata & \nodata & \nodata & & $< 14.9$ & \nodata & \nodata & & $< 11.4$ & \nodata & \nodata \\
\ion{Na}{1} & 5889.95 & 0.000 & $  0.108$ & $ 198.2$ &  $ 0.7$ &  $ 5.7$ & & $ 199.2$ &  $ 2.4$ &  $ 3.5$ & & $ 206.4$ &  $ 3.8$ & \nodata & & $ 144.1$ &  $17.9$ &  $ 5.2$ \\
\ion{Na}{1} & 5895.92 & 0.000 & $ -0.194$ & $ 164.0$ &  $ 1.3$ &  $ 3.9$ & & $ 175.2$ &  $ 1.6$ &  $ 5.3$ & & $ 184.0$ &  $ 5.3$ & \nodata & & $ 106.2$ &  $18.4$ &  $ 3.2$ \\
\ion{Mg}{1} & 3829.36 & 2.710 & $ -0.227$ &  \nodata & \nodata & \nodata & & $ 192.3$ &  $ 0.4$ & \nodata & &  \nodata & \nodata & \nodata & &  \nodata & \nodata & \nodata \\
\ion{Mg}{1} & 4057.51 & 4.350 & $ -0.900$ & $  26.7$ &  $ 0.4$ &  $ 0.7$ & & $  16.7$ &  $ 0.2$ & \nodata & & $  87.0$ &  $10.4$ & \nodata & &  \nodata & \nodata & \nodata \\
\ion{Mg}{1} & 4167.27 & 4.350 & $ -0.745$ & $  36.2$ &  $ 0.6$ &  $ 1.8$ & &  \nodata & \nodata & \nodata & & $ 130.0$ &  $17.4$ &  $12.9$ & &  \nodata & \nodata & \nodata \\
\ion{Mg}{1} & 4702.99 & 4.350 & $ -0.440$ & $  54.6$ &  $ 0.7$ &  $ 1.0$ & &  \nodata & \nodata & \nodata & & $ 115.9$ &  $ 4.7$ &  $ 4.1$ & & $  42.5$ &  $ 4.7$ &  $ 3.6$ \\
\ion{Mg}{1} & 5172.69 & 2.710 & $ -0.393$ & $ 181.1$ &  $ 0.5$ &  $ 5.2$ & & $ 209.7$ &  $ 1.1$ &  $ 5.7$ & & $ 332.3$ &  $ 4.6$ &  $15.5$ & & $ 167.4$ &  $ 4.5$ &  $ 5.2$ \\
\enddata
\tablecomments{(This table is available in its entirety in a machine-readable form in the online journal. A portion is shown here for guidance regarding its form and content.)}
\end{deluxetable}

\setcounter{table}{3}
\begin{deluxetable}{lrcccccrcccccrcccccrcccc}
\tabletypesize{\scriptsize}
\tablecolumns{24}
\tablewidth{0pt}
\tablecaption{Abundances\label{tab:abund}}
\tablehead{ & \multicolumn{5}{c}{HD 115444} & & \multicolumn{5}{c}{HD 122563} & & \multicolumn{5}{c}{Scl 1019417} & & \multicolumn{5}{c}{UMi 20103} \\ \cline{2-6} \cline{8-12} \cline{14-18} \cline{20-24}
\colhead{Element} & \colhead{$N$} & \colhead{$\epsilon$} & \colhead{[X/Fe]} & \colhead{$\delta_{\rm{noise}}\epsilon$} & \colhead{$\delta_{\rm{sys}}\epsilon$} & & \colhead{$N$} & \colhead{$\epsilon$} & \colhead{[X/Fe]} & \colhead{$\delta_{\rm{noise}}\epsilon$} & \colhead{$\delta_{\rm{sys}}\epsilon$} & & \colhead{$N$} & \colhead{$\epsilon$} & \colhead{[X/Fe]} & \colhead{$\delta_{\rm{noise}}\epsilon$} & \colhead{$\delta_{\rm{sys}}\epsilon$} & & \colhead{$N$} & \colhead{$\epsilon$} & \colhead{[X/Fe]} & \colhead{$\delta_{\rm{noise}}\epsilon$} & \colhead{$\delta_{\rm{sys}}\epsilon$}}
\startdata
\protect\ion{Fe}{1}         &  82 & \phs$  4.39$ & $ -3.13$\tablenotemark{a} &  $0.01$ &  $0.01$ & &  49 & \phs$  4.64$ & $ -2.88$\tablenotemark{a} &  $0.01$ &  $0.02$ & &  91 & \phs$  5.12$ & $ -2.40$\tablenotemark{a} &  $0.05$ &  $0.02$ & &  53 & \phs$  4.36$ & $ -3.16$\tablenotemark{a} &  $0.10$ &  $0.02$ \\
\protect\ion{Fe}{2}         &  16 & \phs$  4.40$ & $ -3.12$\tablenotemark{a} &  $0.00$ &  $0.03$ & &  14 & \phs$  4.65$ & $ -2.87$\tablenotemark{a} &  $0.00$ &  $0.04$ & &  16 & \phs$  4.97$ & $ -2.55$\tablenotemark{a} &  $0.03$ &  $0.04$ & &   8 & \phs$  4.52$ & $ -3.00$\tablenotemark{a} &  $0.06$ &  $0.07$ \\
\protect\ion{Li}{1}         &   0 &   \nodata    & \nodata  & \nodata & \nodata & &   0 &   \nodata    & \nodata  & \nodata & \nodata & &   1 &     $<-0.24$ & $<-1.15$ & \nodata & \nodata & &   1 & \phs$< 0.86$ & $<+0.71$ & \nodata & \nodata \\
        C (CH)              &   syn & \phs$ 5.26$ & $-0.17$ & $0.00$ & $0.20$ & &   syn & \phs$ 5.27$ & $-0.41$ & $0.00$ & $0.20$ & &   syn & \phs$ 5.19$ & $-0.97$ & $0.01$ & $0.20$ & &   syn & \phs$ 4.54$ & $-0.85$ & $0.05$ & $0.20$ \\
\protect\ion{O}{1}          &   1 & \phs$  6.56$ & $ +0.76$ &  $0.04$ & \nodata & &   0 &   \nodata    & \nodata  & \nodata & \nodata & &   1 & \phs$< 7.26$ & $<+0.73$ & \nodata & \nodata & &   1 & \phs$< 7.75$ & $<+1.98$ & \nodata & \nodata \\
\protect\ion{Na}{1}         &   2 & \phs$  3.64$ & $ +0.44$ &  $0.03$ &  $0.06$ & &   2 & \phs$  3.45$ & $ -0.00$ &  $0.02$ &  $0.01$ & &   2 & \phs$  3.17$ & $ -0.76$ &  $0.10$ &  $0.02$ & &   2 & \phs$  2.95$ & $ -0.22$ &  $0.22$ &  $0.10$ \\
\protect\ion{Mg}{1}         &   7 & \phs$  5.00$ & $ +0.55$ &  $0.01$ &  $0.01$ & &   6 & \phs$  4.98$ & $ +0.27$ &  $0.01$ &  $0.11$ & &   7 & \phs$  5.74$ & $ +0.57$ &  $0.04$ &  $0.07$ & &   4 & \phs$  4.81$ & $ +0.39$ &  $0.09$ &  $0.01$ \\
\protect\ion{Al}{1}         &   1 & \phs$  3.13$ & $ -0.21$ &  $0.03$ & \nodata & &   1 & \phs$  3.32$ & $ -0.27$ &  $0.02$ & \nodata & &   1 & \phs$  3.70$ & $ -0.37$ &  $0.39$ & \nodata & &   1 & \phs$  2.66$ & $ -0.65$ &  $0.41$ & \nodata \\
\protect\ion{Si}{1}         &   4 & \phs$  5.06$ & $ +0.64$ &  $0.01$ &  $0.04$ & &   1 & \phs$  5.21$ & $ +0.54$ &  $0.01$ & \nodata & &   6 & \phs$  5.67$ & $ +0.53$ &  $0.02$ &  $0.04$ & &   1 & \phs$  4.77$ & $ +0.38$ &  $0.21$ & \nodata \\
\protect\ion{K}{1}          &   1 & \phs$  2.30$ & $ +0.31$ &  $0.01$ & \nodata & &   0 &   \nodata    & \nodata  & \nodata & \nodata & &   1 & \phs$  3.05$ & $ +0.33$ &  $0.07$ & \nodata & &   1 & \phs$< 2.65$ & $<+0.69$ & \nodata & \nodata \\
\protect\ion{Ca}{1}         &  19 & \phs$  3.63$ & $ +0.40$ &  $0.01$ &  $0.03$ & &  13 & \phs$  3.81$ & $ +0.33$ &  $0.01$ &  $0.05$ & &  22 & \phs$  4.15$ & $ +0.19$ &  $0.04$ &  $0.04$ & &   6 & \phs$  3.42$ & $ +0.22$ &  $0.08$ &  $0.06$ \\
\protect\ion{Sc}{2}         &   9 & \phs$  0.17$ & $ +0.19$ &  $0.01$ &  $0.04$ & &   6 & \phs$  0.47$ & $ +0.24$ &  $0.00$ &  $0.07$ & &  11 & \phs$  0.98$ & $ +0.43$ &  $0.03$ &  $0.03$ & &   4 & \phs$  0.14$ & $ +0.03$ &  $0.09$ &  $0.03$ \\
\protect\ion{Ti}{1}         &  19 & \phs$  2.25$ & $ +0.39$ &  $0.01$ &  $0.01$ & &  15 & \phs$  2.37$ & $ +0.25$ &  $0.01$ &  $0.02$ & &  17 & \phs$  2.67$ & $ +0.08$ &  $0.07$ &  $0.03$ & &   4 & \phs$  2.07$ & $ +0.25$ &  $0.10$ &  $0.05$ \\
\protect\ion{Ti}{2}         &  22 & \phs$  2.38$ & $ +0.51$ &  $0.01$ &  $0.03$ & &  18 & \phs$  2.53$ & $ +0.41$ &  $0.01$ &  $0.04$ & &  23 & \phs$  2.84$ & $ +0.40$ &  $0.04$ &  $0.05$ & &  14 & \phs$  2.08$ & $ +0.08$ &  $0.08$ &  $0.06$ \\
\protect\ion{V}{1}          &   1 & \phs$  0.86$ & $ -0.01$ &  $0.04$ & \nodata & &   0 &   \nodata    & \nodata  & \nodata & \nodata & &   2 & \phs$  1.56$ & $ -0.03$ &  $0.09$ &  $0.29$ & &   1 & \phs$< 2.18$ & $<+1.34$ & \nodata & \nodata \\
\protect\ion{V}{2}          &   0 &   \nodata    & \nodata  & \nodata & \nodata & &   2 & \phs$  1.04$ & $ -0.09$ &  $0.01$ &  $0.05$ & &   0 &   \nodata    & \nodata  & \nodata & \nodata & &   0 &   \nodata    & \nodata  & \nodata & \nodata \\
\protect\ion{Cr}{1}         &  11 & \phs$  2.32$ & $ -0.22$ &  $0.01$ &  $0.04$ & &  14 & \phs$  2.56$ & $ -0.24$ &  $0.01$ &  $0.05$ & &  12 & \phs$  2.98$ & $ -0.29$ &  $0.07$ &  $0.07$ & &   5 & \phs$  2.02$ & $ -0.49$ &  $0.15$ &  $0.10$ \\
\protect\ion{Mn}{1}         &   5 & \phs$  1.70$ & $ -0.56$ &  $0.01$ &  $0.12$ & &   6 & \phs$  1.95$ & $ -0.56$ &  $0.01$ &  $0.16$ & &   7 & \phs$  2.55$ & $ -0.44$ &  $0.06$ &  $0.04$ & &   2 & \phs$  1.31$ & $ -0.91$ &  $0.23$ &  $0.19$ \\
\protect\ion{Co}{1}         &   2 & \phs$  1.91$ & $ +0.12$ &  $0.01$ &  $0.19$ & &   3 & \phs$  2.17$ & $ +0.12$ &  $0.01$ &  $0.14$ & &   3 & \phs$  2.58$ & $ +0.07$ &  $0.06$ &  $0.23$ & &   1 & \phs$  1.82$ & $ +0.06$ &  $0.13$ & \nodata \\
\protect\ion{Ni}{1}         &   5 & \phs$  3.11$ & $ -0.00$ &  $0.01$ &  $0.09$ & &   7 & \phs$  3.49$ & $ +0.12$ &  $0.01$ &  $0.11$ & &  18 & \phs$  3.72$ & $ -0.12$ &  $0.04$ &  $0.05$ & &   1 & \phs$< 3.35$ & $<+0.26$ & \nodata & \nodata \\
\protect\ion{Cu}{1}         &   0 &   \nodata    & \nodata  & \nodata & \nodata & &   0 &   \nodata    & \nodata  & \nodata & \nodata & &   1 & \phs$  1.19$ & $ -0.62$ &  $0.06$ & \nodata & &   1 & \phs$< 1.82$ & $<+0.77$ & \nodata & \nodata \\
\protect\ion{Zn}{1}         &   2 & \phs$  1.79$ & $ +0.32$ &  $0.01$ &  $0.03$ & &   0 &   \nodata    & \nodata  & \nodata & \nodata & &   2 & \phs$  2.18$ & $ -0.02$ &  $0.06$ &  $0.11$ & &   1 & \phs$< 2.19$ & $<+0.75$ & \nodata & \nodata \\
\protect\ion{Sr}{2}         &   2 & \phs$  0.02$ & $ +0.23$ &  $0.04$ &  $0.04$ & &   0 &   \nodata    & \nodata  & \nodata & \nodata & &   2 & \phs$  1.69$ & $ +1.34$ &  $0.36$ &  $0.13$ & &   2 &     $ -0.62$ & $ -0.52$ &  $0.28$ &  $0.07$ \\
\protect\ion{Y}{2}          &   8 &     $ -0.92$ & $ -0.05$ &  $0.01$ &  $0.02$ & &  10 &     $ -0.85$ & $ -0.21$ &  $0.01$ &  $0.04$ & &   7 &     $ -0.25$ & $ +0.06$ &  $0.04$ &  $0.03$ & &   1 &     $<-1.07$ & $<-0.31$ & \nodata & \nodata \\
\protect\ion{Zr}{2}         &   5 &     $ -0.03$ & $ +0.48$ &  $0.01$ &  $0.05$ & &   2 &     $ -0.34$ & $ -0.06$ &  $0.01$ &  $0.26$ & &   4 & \phs$  0.77$ & $ +0.72$ &  $0.09$ &  $0.09$ & &   1 & \phs$< 0.06$ & $<+0.46$ & \nodata & \nodata \\
\protect\ion{Ba}{2}         &   5 &     $ -0.75$ & $ +0.23$ &  $0.01$ &  $0.02$ & &   3 &     $ -1.62$ & $ -0.87$ &  $0.01$ &  $0.15$ & &   5 &     $ -0.86$ & $ -0.44$ &  $0.05$ &  $0.05$ & &   4 &     $ -1.80$ & $ -0.93$ &  $0.08$ &  $0.10$ \\
\protect\ion{La}{2}         &   7 &     $ -1.58$ & $ +0.32$ &  $0.01$ &  $0.02$ & &   0 &   \nodata    & \nodata  & \nodata & \nodata & &   1 &     $<-2.05$ & $<-0.72$ & \nodata & \nodata & &   1 &     $<-1.46$ & $<+0.32$ & \nodata & \nodata \\
\protect\ion{Ce}{2}         &   5 &     $ -1.20$ & $ +0.36$ &  $0.01$ &  $0.06$ & &   0 &   \nodata    & \nodata  & \nodata & \nodata & &   1 &     $<-1.43$ & $<-0.43$ & \nodata & \nodata & &   1 &     $<-0.85$ & $<+0.60$ & \nodata & \nodata \\
\protect\ion{Nd}{2}         &  10 &     $ -1.00$ & $ +0.62$ &  $0.01$ &  $0.04$ & &   1 &     $ -1.86$ & $ -0.49$ &  $0.02$ & \nodata & &   1 &     $<-1.37$ & $<-0.32$ & \nodata & \nodata & &   1 &     $<-1.10$ & $<+0.40$ & \nodata & \nodata \\
\protect\ion{Sm}{2}         &   4 &     $ -1.41$ & $ +0.70$ &  $0.02$ &  $0.11$ & &   0 &   \nodata    & \nodata  & \nodata & \nodata & &   1 &     $<-1.44$ & $<+0.11$ & \nodata & \nodata & &   1 &     $<-0.96$ & $<+1.04$ & \nodata & \nodata \\
\protect\ion{Eu}{2}         &   3 &     $ -1.82$ & $ +0.79$ &  $0.01$ &  $0.13$ & &   0 &   \nodata    & \nodata  & \nodata & \nodata & &   1 &     $ -1.65$ & $ +0.39$ &  $0.14$ & \nodata & &   1 &     $<-2.26$ & $<+0.23$ & \nodata & \nodata \\
\protect\ion{Gd}{2}         &   0 &   \nodata    & \nodata  & \nodata & \nodata & &   0 &   \nodata    & \nodata  & \nodata & \nodata & &   1 &     $<-0.43$ & $<+1.00$ & \nodata & \nodata & &   1 &     $<-0.28$ & $<+1.60$ & \nodata & \nodata \\
\protect\ion{Dy}{2}         &   2 &     $ -1.11$ & $ +0.91$ &  $0.01$ &  $0.09$ & &   1 &     $ -2.01$ & $ -0.24$ &  $0.02$ & \nodata & &   1 &     $<-1.37$ & $<+0.08$ & \nodata & \nodata & &   1 & \phs$< 0.18$ & $<+2.08$ & \nodata & \nodata \\
\protect\ion{Ho}{2}         &   0 &   \nodata    & \nodata  & \nodata & \nodata & &   1 &     $ -1.93$ & $ +0.44$ &  $0.01$ & \nodata & &   0 &   \nodata    & \nodata  & \nodata & \nodata & &   0 &   \nodata    & \nodata  & \nodata & \nodata \\
\protect\ion{Pb}{1}         &   0 &   \nodata    & \nodata  & \nodata & \nodata & &   0 &   \nodata    & \nodata  & \nodata & \nodata & &   1 & \phs$< 0.34$ & $<+0.89$ & \nodata & \nodata & &   1 & \phs$< 0.98$ & $<+2.29$ & \nodata & \nodata \\
\enddata
\tablecomments{NLTE corrections have been applied to \ion{Na}{1}, \ion{Al}{1}, and \ion{K}{1}.}
\tablenotetext{a}{[Fe/H]}
\end{deluxetable}
\clearpage
\end{turnpage}

\end{document}